\documentclass[twocolumn,amsmath,showpacs,amssymb, floatfix, nofootinbib,10pt]{revtex4}

\usepackage{indentfirst}
\usepackage{tikz}
\usepackage{setspace}
\usepackage{amsfonts,color,xcolor,graphicx,amsfonts,multirow,amsmath}
\usepackage[section]{placeins}
\definecolor{bluee}{rgb}{0,0,1}
\usepackage[colorlinks=true, colorlinks=true, citecolor=bluee, linkcolor=bluee, urlcolor=bluee]{hyperref}

\usepackage{txfonts}
\DeclareMathAlphabet{\mathcal}{OMS}{cmsy}{m}{n}
\DeclareSymbolFont{largesymbols}{OMX}{cmex}{m}{n}
\definecolor{darkerblue}{rgb}{0.1, 0.1, 0.7}
\definecolor{darkergreen}{rgb}{0.1, 0.5, 0.1}
\definecolor{lightred}{rgb}{1, 0, 0}

\usepackage{ulem}

\allowdisplaybreaks   

\usepackage{slashed}   
\usepackage{cancel}

\begin{document}
\title{Analysis of $S$-wave $\omega\pi$ dynamics behavior in multi-body $B_{\ell4}$ and $B_{\ell6}$ decays}

\author{Yan-Ting Yang}
\author{Hai-Bing Fu}
\email{fuhb@gzmu.edu.cn (corresponding author)}
\author{Xue Zheng}
\author{Jian-Qi Chen}
\author{Sheng-Bo Wu}
\address{Department of Physics, Guizhou Minzu University, Guiyang 550025, P.R.China}

\begin{abstract}
The $b_1(1235)$ state has a sizable decay width, which have longstanding puzzle in treating as a final state in theoretical calculations may lead to systematic deviations in the predicted observables. To address this issue and further explore the internal structure of this resonance, we study the multi-body semileptonic decays $B_{\ell4}$ and $B_{\ell6}$. Within the approximation where the $b_1(1235)$ resonance dominates, we study the $B_{\ell4}$ decay $B\to b_1(1235)(\to\omega\pi)\ell^+\nu_\ell$ in the QCD light-cone sum rules framework. The corresponding transition form factors (TFFs) are calculated by constructing a correlation function with a right-handed chiral current. As the main nonperturbative input for the TFFs calculation, the transverse twist-2 light-cone distribution amplitude (LCDA) of the $b_1(1235)$ resonance is constructed using the light-cone harmonic oscillator model, and the resulting LCDA exhibits a single-peak behavior. At the large-recoil region, we obtain the TFFs $V_{0, 1, 2}(0)$ and $A(0)$, together with the corresponding TFFs ratios $r_V = 1.530_{-0.016}^{+0.013}$ and $r_2 = 0.553_{-0.015}^{+0.012}$. We then further study the complete $B_{\ell4}$ decay by incorporating the $b_1(1235)\to\omega\pi$ resonant line shape and obtain the corresponding observables. The results show that the branching ratio of the $B_{\ell4}$ decay is about $1/3$ of that obtained when $b_1(1235)$ is treated as a final state. We also account for the subsequent decay $\omega \to \pi^+\pi^-\pi^0$ and further predicted the branching ratio of the $B_{\ell6}$ decay $B\to b_1(1235)(\to\omega(\to \pi^+\pi^-\pi^0)\pi)\ell^+\nu_\ell$. These results may provide useful references for future experimental measurements and further theoretical studies.
\end{abstract}
\date{\today}
\maketitle

\section{Introduction}

In the low-energy region of strong interactions, information about the internal structure of excited hadronic states is often difficult to extract directly from their complex invariant-mass and angular distributions. This is particularly relevant for axial-vector states, which may have rich internal configurations and sizable finite-width effects. Identifying these effects through experimentally accessible observables is therefore important for understanding the structure of light hadrons. Semileptonic decays of heavy-flavor $B$-mesons provide a useful platform for studying weak transitions and hadronic transition dynamics~\cite{ParticleDataGroup:2026mpi}. In such processes, the resonance of interest can be produced through a relatively clean short-distance weak transition, while its properties are subsequently encoded in the experimentally measurable multi-body final states through its strong decay.

At present, several axial-vector states are known to have sizable decay widths. Both experimental and theoretical studies of these states therefore often rely on the multi-body final states formed by their decay products. Semileptonic decays provide a relatively clean environment for such investigations. Since the final-state lepton does not participate in strong interactions, the hadronic dynamics is mainly governed by the weak-current-induced transition and the subsequent strong interactions among the final-state hadrons. Compared with purely hadronic decays, semileptonic processes generally suffer from reduced hadronic backgrounds and are therefore better suited for studying the spectroscopic properties and internal structure of axial-vector resonances~\cite{Cheng:2017smj,Yang:2005gk}. Experimentally, in 2024, the BESIII Collaboration reported the first observation of $D^0\to b_1(1235)^-e^+\nu_e$ with a statistical significance of $5.2\sigma$, and also found evidence for $D^+\to b_1(1235)^0e^+\nu_e$ with a significance of $3.1\sigma$~\cite{BESIII:2024pwp}. In both decay channels, the $b_1(1235)$ resonance was reconstructed through the $\omega\pi$ final state. These measurements provide direct experimental motivation for further studies of semileptonic decays involving the $b_1(1235)$ resonance. On the theoretical side, On the theoretical side, the $B_{\ell4}$ decays $D\to M(M\to VP)\ell^+\nu_\ell$ were systematically studied within SU$_f$(3) symmetry, where $M = A/T$ and $A$, $T$, $V$, and $P$ denote axial-vector, tensor, vector, and pseudoscalar states, respectively~\cite{Qiao:2024nbq}. It was shown that both axial-vector and tensor resonances can contribute to such decay modes, indicating that multi-body semileptonic decays provide a useful approach for studying resonance structures in hadronic final states. In addition, lattice QCD studies indicate a strong coupling of the $b_1(1235)$ resonance to the $S$-wave $\omega\pi$ channel~\cite{Lang:2014tia}, further supporting studies of this resonance through the $\omega\pi$ final state. According to the PDG~\cite{ParticleDataGroup:2026mpi}, the $\omega$-meson decays predominantly through $\omega\to\pi^+\pi^-\pi^0$, with a branching fraction of $\mathcal{B}(\omega\to\pi^+\pi^-\pi^0) = 89.2\pm0.7\%$. Moreover, its natural width, $\Gamma_\omega = 8.68\pm0.13~{\rm MeV}$, is much smaller than that of the $b_1(1235)$ resonance, $\Gamma_{b_1} = 142\pm9~{\rm MeV}$. Therefore, the $b_1(1235)$ resonance can be experimentally reconstructed through the corresponding four-pion final state, and the complete decay chain can be written as $B\to b_1(1235)(\to\omega(\to\pi^+\pi^-\pi^0)\pi)\ell^+\nu_\ell$. Despite these experimental and theoretical developments, systematic studies of the $B_{\ell4}$ decay $B\to b_1(1235)(\to\omega\pi)\ell^+\nu_\ell$ remain limited. A detailed investigation of this process is therefore worthwhile for studying the $B\to b_1(1235)$ transition dynamics, the finite-width effect of the $b_1(1235)$ resonance, and the resonance structure of the $\omega\pi$ final state.

Axial-vector states possess a rich internal structure. They have the same total angular momentum and parity, but different states show different properties because of differences in symmetry, composition, and dominant decay channel. For non-strange systems, $b_1(1235)$ and $h_1(1170)$ are classified as $1^{+-}$-states, whereas $a_1(1260)$, $f_1(1285)$ and $f_1(1420)$ are classified as $1^{++}$-states. The strange states $K_1(1270)$ and $K_1(1400)$ have no definite $C$ or $G$-parity, they involve mixing between the $^1P_1$ and $^3P_1$ basis states. According to the latest PDG compilation~\cite{ParticleDataGroup:2026mpi}, The mass of the $b_1(1235)$ state is $m_{b_1(1235)} = 1229.5\pm3.2~\rm MeV$, its average width is $\Gamma_{b_1(1235)} = 142\pm9~\rm MeV$ and its main decay channels are $\omega\pi$~\cite{Chen:2015iqa}. Which compared with other axial-vector states, The mass and width of $a_1(1260)$ are $375\pm35~\rm MeV$ and $250--600~{\rm MeV}$, respectively. The dominant decay channel for both $h_1(1170)$ and $a_1(1260)$ is $\rho\pi$\cite{Dankowych:1981ks,OmegaPhoton:1983huz,Ando:1990ti,CLEO:1999rzk}, and both states have relatively large decay widths. The widths of $f_1(1285)$ ($23.0\pm1.1~\rm MeV$) and $f_1(1420)$ ($56.7\pm 3.3~\rm MeV$) are relatively narrow, and they mainly decay into $\eta\pi\pi$~\cite{Dorofeev:2011zz, Corden:1978cz, Amsterdam-CERN-Nijmegen-Oxford:1978mws} and $K\bar{K}\pi$~\cite{Dorofeev:2011zz, Corden:1978cz, WA102:1998zhh, Bromberg:1980bk, CERN-CollegedeFrance-Madrid-Stockholm:1980umk}. Although $b_1(1235)$ and $a_1(1260)$ have similar masses, they differ in their internal quantum numbers and $G$-parities. The considerable width of $b_1(1235)$ resonance makes it an important tool for studying the multi-body dynamics of axial-vector resonances.

Currently, the internal structure of axial-vector states is still not fully understood.  In particular, the internal structure of $b_1(1235)$ state has not yet been completely determined. The $b_1(1235)$ state was first observed by the HBC in 1963~\cite{Abolins:1963zz}. Before 1994, studies on the $b_1(1235)$ resonance mainly focused on processes such as $p\bar{p}$~\cite{Bizzarri:1969snf, Baltay:1967zza}, $\gamma p$~\cite{OmegaPhoton:1984ols}, and $p\pi$~\cite{Karshon:1974qi, Fayet:1974jb, Gessaroli:1977mt, Evangelista:1980xe, OmegaPhoton:1983vde, Collick:1984dkp, Fukui:1990ki, IHEP-IISN-LANL-LAPP-KEK:1992puu, ASTERIX:1993wam}, with the main goal of measuring its mass and width. In the processes $p\bar{p} \to \omega^0\pi^+\pi^-$ at rest, the $b_1(1235)$ (then called the $B$) was clearly observed. Its mass, width, and spin-parity were determined to be approximately $1200~{\rm MeV}$, $100-130~{\rm MeV}$, and $J^{P} = 1^+$, respectively~\cite{Bizzarri:1969snf, Baltay:1967zza}. For the $\pi^+ p $ collision experiment in 1974, the $\omega\pi$ decay channel was analyzed at $5 ~\rm GeV/c$, and the mass and width of the $b_1(1235)$ resonance were measured, yielding values similar to those obtained in $p\bar{p} \to \omega^0\pi^+\pi^-$~\cite{Karshon:1974qi,Gessaroli:1977mt}. It is indicated by these results that the $b_1(1235)$ is a resonance with a significant width. In the traditional quark model, $b_1(1235)$ is often described as a quark-antiquark pair in the $^1P_1$ state~\cite{Godfrey:1985xj,Ebert:2009ub}. Lattice QCD calculations show that the $b_1(1235)$ resonance can be effectively produced by an operator composed of a quark and an antiquark. This indicates that it has a significant $q\bar{q}$ component~\cite{Jansen:2009yh}. The calculations also show that the interaction between $b_1(1235)$ and the $\omega\pi$ system is also important~\cite{Lang:2014tia}. Some theoretical studies have proposed different explanations. For example, in the diquark-antidiquark color-spin model, $b_1(1235)$ can be regarded as a candidate state of the $J^{PC} = 1^{+-}$ tetraquark state~\cite{Kim:2016dfq}. Based on the above discussion, more theoretical support is given to $b_1(1235)$ as a $q\bar{q}$ state, and the subsequent study in this paper is carried out on this basis.
\begin{table}[t]

\footnotesize
\begin{center}
\caption{The LCDAs of vector and axial-vector states exhibit symmetric or antisymmetric behavior under the exchange of quark and antiquark momentum fractions~\cite{Yang:2007zt}.}
\label{table:Cpara}
\begin{tabular}{l @{\hspace{2em}}l@{\hspace{2em}}ll}
\hline
&$\phi_{2}^{\|}$ &$\phi_{2}^\bot$  \\
\hline
Vector state & symmetric & symmetric \\
Axial-vector state $(1^3P_1)$   & symmetric & antisymmetric \\
Axial-vector state $(1^1P_1)$   & antisymmetric & symmetric  \\
\hline
\end{tabular}
\end{center}
\end{table}

For the $B_{\ell4}$ decay $B\to b_1(1235)(\to\omega\pi)\ell^+\nu_\ell$, the decay dynamics involves both the weak transition $B\to b_1(1235)$ and the subsequent strong decay $b_1(1235)\to\omega\pi$. These two parts are connected through the $b_1(1235)$ propagator. To describe the resonance line shape and the finite-width effect near the $\omega\pi$ threshold, we adopt a Breit-Wigner form with a mass-dependent width~\cite{BaBar:2010vmf}. The nonperturbative dynamics of the weak transition is encoded in the $B\to b_1(1235)$ transition form factors (TFFs), which therefore serve as key hadronic inputs for calculating the differential decay rates and other physical observables. A reliable determination of these TFFs is essential for a quantitative description of the four-body decay. In particular, precise knowledge of the $B\to b_1(1235)$ TFFs provides an important theoretical basis for studying various observables of this $B_{\ell4}$ decay and for assessing the finite-width effect of the $b_1(1235)$ resonance.

Theoretically, many methods are available for studying heavy $B$-meson to the light axial-vector $b_1(1235)$ TFFs, including perturbative QCD (pQCD)~\cite{Li:2009tx}, QCD light-cone sum rules (LCSRs)~\cite{Sun:2011ssd,Yang:2008xw}, and the covariant light-front quark model (CLFQM)~\cite{Kang:2018jzg,Cheng:2003sm}. For pQCD approach in 2009, the factorization scheme with retains the transverse momentum $k_T$ is adopted where the TFFs were calculated TFFs at large-recoil region and whole physical region directly. Branching ratios for $B$-meson decays to different axial-vector states is also given, most of them are at the $10^{-4}$ level. It was also pointed out that the sizes of the TFFs and different helicity components are very sensitive to the distribution amplitudes of the axial-vector state~\cite{Li:2009tx}. Implementation of the CLFQM method, a scheme was adopted in which the calculations were carried out in the spacelike region and then extended to the timelike region through a parametrization. Within this scheme, the TFFs were treated separately for the spacelike and timelike regions. It was also pointed out that the internal structure of axial-vector states could significantly affect the related TFFs predictions~\cite{Cheng:2003sm}. For the LCSR method in 2008, the vector and axial-vector TFFs were calculated for $B_{u,d,s}$ decays to $P$-wave axial-vector states, with $m_A/m_b$ order corrections included. The differential decay widths and branching ratios for the subsequent $B_{u,d,s} \to A e \nu_e$ were measured~\cite{Yang:2008xw}. A basis for evaluating the reliability of TFFs predictions can be provided by comparing different results~\cite{Melikhov:2000yu}. In this work, we use the LCSR method, it is suitable for the low to mid $q^2$-region required for this process, the LCDA of $b_1(1235)$ resonance can be used to describe the relevant nonperturbative effects. The uncertainties of nonperturbative input can also be evaluated. This method was developed from 1980s. It can be regarded as an extension of the SVZ sum rules to the light cone~\cite{Balitsky:1989ry,Chernyak:1990ag}.

To accurately determine the behavior and precision of the TFFs in the $B\to b_1(1235)$ transition, the leading-twist LCDAs of the $b_1(1235)$ state play a key role, which contain information about the momentum distribution of quarks inside the hadron, Therefore, we also pay close attention to the behavior of its twist-2 LCDA amplitude. Within the LCSR framework, for axial-vector state, the LCDAs include both longitudinal and transverse polarization states, which are described by chiral-even and chiral-odd operators, respectively. Regarding LCDAs, opposite symmetry properties are exhibited by the chiral-even and chiral-odd cases under quark-antiquark momentum exchange~\cite{Yang:2007zt}. From the QCDSR~\cite{Yang:2007zt} sum rules, it can be known that the chiral-even two-particle LCDAs of axial-vector state are antisymmetric, while the opposite holds for the chiral-odd LCDAs. The symmetric and antisymmetric properties of the twist-2 LCDAs of axial-vector states under the exchange of quark momentum fractions are listed in Table~\ref{table:Cpara}. For the longitudinally polarized one couples through the axial-vector current $\bar{u}\gamma _{\mu}\gamma _5b$, while the transversely polarized $b_1(1235)$ state is described by the tensor current $\bar{u}\sigma_{\mu\nu}b$. We use the right-handed chiral current to construct the correlation function and obtain the TFFs. We choose the chiral current over the conventional current to construct the correlation function. Within the present QCD light-cone sum rule framework, this choice helps suppress unwanted resonance contributions, allows some strong interaction corrections to cancel, and reduces the sensitivity of the results to higher-twist distribution amplitudes~\cite{Huang:1998gp, Huang:2001xb, Wang:2002aje, Fu:2014pba}, These features can improve the stability of the calculation and the precision of theoretical predictions~\cite{Wang:2002aje, Fu:2014pba}. For the transverse leading-twist distribution amplitude $\phi _{2;b_1(1235)}^\bot(x,\mu)$, which gives the dominant contribution, the light-cone harmonic oscillator (LCHO) model is adopted to construct it. This model is based on the Brodsky-Huang-Lepage (BHL) prescription, which connects the equal-time wave function of the meson in its rest frame to the light-cone wave function in the infinite momentum frame~\cite{Zhong:2011rg,Wu:2007rt,Huang:2004su}, it provides an intuitive and effective framework for modeling the internal structure of mesons. This model has been successfully applied to pseudoscalar mesons $\pi$, $K$, $\eta ^{(\prime)}$~\cite{Zhong:2022ecl, Zhong:2021epq, Hu:2023pdl}, scalar states $a_0(980)$, $K_{0}^*\left( 1430 \right)$~\cite{Huang:2022xny,Yang:2024ang,Wu:2022qqx} and vector states $\rho $, $K^*$, $\phi$~\cite{Hu:2024tmc,Fu:2016yzx,Fu:2014uea}. At the same time, for axial-vector states, the LCHO model has also been successfully used to construct the longitudinal twist-2 LCDA of $a_1(1260)$. It has also been applied to study the $D \to a_1(1260)$ TFFs and semileptonic decays~\cite{Hu:2021lkl}. This shows that the LCHO model can provide an effective parametrization for the nonperturbative structure of axial-vector states. Based on this, in this work, we adopt the LCHO model to construct the transverse twist-2 LCDA $\phi _{2;b_1(1235)}^\bot(x,\mu)$ of $b_1(1235)$ as an input parameter. Then, we use LCSR to study the semileptonic decay process with a four-body final state.

The structure of this paper is organized as follows: In Section ~\ref{sec:framework}, the decay width of the $B_{\ell4}$ decay $B\to b_1(1235)(\to\omega\pi)\ell^+ \nu_\ell$ and the theoretical framework for the calculation of the TFFs by means of the LCSR method are first briefly introduced. This includes the detailed procedure for the construction of $\phi _{2;b_1(1235)}^\bot(x,\mu)$ in the LCHO model, the method by which the correlation functions are built with the right-handed chiral current, and the expressions for the TFFs. Section ~\ref{sec:analysis} presents the numerical results and analysis, covering the behavior of the LCDAs, the shapes of the TFFs, the differential decay width, the branching fraction, Finally, the paper concludes with a summary and outlook.

\section{Theoretical Framework}\label{sec:framework}
\begin{figure}[t]
\begin{center}
\includegraphics[width = 0.42\textwidth]{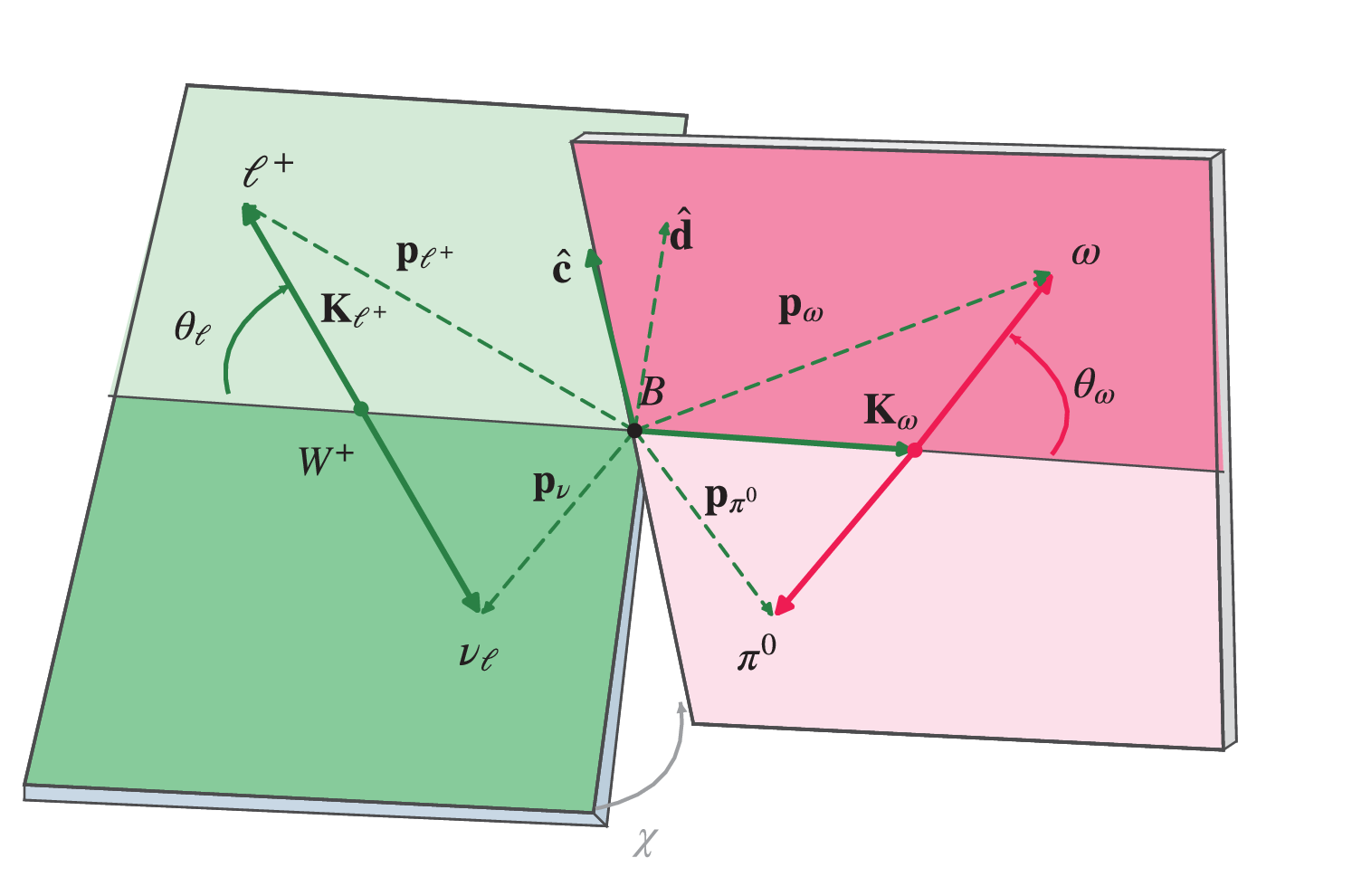}
\caption{Definition of angular variable, among them, the three-momenta of $\ell^+$ in the rest frame of the virtual $W^+$ and of $\omega\pi$ rest frame are denoted by $K_{\ell^+}$ and $K_{\omega}$, respectively. The unit vector along the $\omega\pi$ direction in the $B$ rest frame is set as $\hat{v}$, and  the unit vectors along the components of ${K}_{\omega}$ and ${K}_{\ell^+}$ that are perpendicular to $\hat{v}$ are defined as $\hat{c}$ and $\hat{d}$, respectively.}
\label{fig:b1omegaJao}
\end{center}
\end{figure}
A general expression for the differential decay width of $B_{\ell4}$ decays has been derived in the Ref.~\cite{Lee:1992ih}. It is formulated in terms of the five-variable Cabibbo-Maksymowicz parametrization~\cite{Cabibbo:1965zzb} and has been widely used in theoretical studies of $B_{\ell4}$ processes. In this work, we adopt this general formalism to describe the decay $B\to b_1(1235)(\to\omega\pi)\ell^+\nu_\ell$. The effective weak interaction amplitude for the $B\to b_1(1235)\ell^+\nu_\ell$ transition can be written as
\begin{align}
{\cal M}_{\rm weak}
 = \frac{G_F}{\sqrt{2}}V_{ub}L^\mu H_\mu ,
\end{align}
where $G_F = 1.166\times10^{-5}~{\rm GeV}^{-2}$ is the Fermi coupling constant. For a positively charged final-state lepton, the leptonic current is defined as
$L^\mu = \bar{u}(p_\nu)\gamma^\mu(1-\gamma_5)v(p_\ell)$, which can be calculated perturbatively. The hadronic current $H_\mu$ describes the charged-current matrix element of the $B\to b_1(1235)$ transition. Its nonperturbative dynamics is parametrized in terms of a finite set of independent TFFs through a Lorentz-covariant decomposition of the matrix element. For the $B_{\ell4}$ decay, the weak production of the $b_1(1235)$ resonance is followed by the strong decay $b_1(1235)\to\omega\pi$. These two parts are connected through the $b_1(1235)$ propagator. The corresponding resonance line shape is described by a Breit-Wigner form with a mass-dependent width, which allows the finite-width effect of the $b_1(1235)$ resonance to be incorporated. The differential decay width is given by $d\Gamma  =  |{\cal M}|^2 d\Phi_4 / (2m_B)$, where $d\Phi_4$ denotes the four-body Lorentz-invariant phase space. For the process $B\to b_1(1235)(\to \omega\pi)\ell^+\nu_\ell$, the phase space can be factorized into the successive two-body phase spaces
$B\to(\omega\pi)W^*,\,(\omega\pi)\to\omega\pi,\,W^*\to\ell^+\nu_\ell$. After integrating over the overall orientation of the spinless initial $B$ meson and neglecting the charged-lepton mass, the final-state kinematics can be completely described by five variables, as shown in Fig.~\ref{fig:b1omegaJao}. The five variables are the invariant mass squared of the $\omega\pi$ system, $s\equiv m_{\omega\pi}^2 = (p_\omega+p_\pi)^2$; the invariant mass squared of the $\ell^+\nu_\ell$ system, $q^2\equiv (p_\ell+p_\nu)^2$; the polar angles $\theta_\omega$ and $\theta_\ell$ defined in the $\omega\pi$ and $\ell^+\nu_\ell$ rest frames, respectively; and the azimuthal angle $\chi$ between the two decay planes. Here, $p_\omega$, $p_\pi$, $p_\ell$ and $p_\nu$ denote the four-momenta of the $\omega$ meson, the pion, the charged lepton, and the neutrino, respectively. In terms of these five variables, the differential decay width can be written as
\begin{align}
\frac{d^5 \Gamma}{ds dq^2 d\cos\theta_\omega d\cos\theta_\ell d\chi}
& =  \frac{G_F^2 |V_{ub}|^2}{(4\pi)^6 m_B^3} X(s,q^2)\beta(s) {\cal I},
\label{sitidT}
\end{align}
where $X(s,q^2) = m_B|p_{\omega\pi}(s,q^2)|$ and $\beta(s) = 2|p^*(s)|/\sqrt{s}$. Here, $|p_{\omega\pi}|$ denotes the magnitude of the three-momentum of the $\omega\pi$ system in the $B$-meson rest frame, while $|p^*(s)|$ denotes the magnitude of the three-momentum of the $\omega$ (or equivalently the $\pi$) in the $\omega\pi$ rest frame. The angular function ${\cal I}$ can be decomposed into nine angular terms, ${\cal I}_{1}$--${\cal I}_{9}$, which depend on $s$, $q^2$, and $\theta_\omega$~\cite{Lee:1992ih}. The physical phase-space region for the $B_{\ell4}$ decay $B\to b_1(1235)(\to\omega\pi)\ell^+\nu_\ell$ is determined by energy-momentum conservation. The allowed kinematic ranges are $(m_\omega + m_\pi)^2 \leq s \leq (m_B - m_\ell)^2$ and $ m_\ell^2 \leq q^2\leq (m_B-\sqrt{s})^2$. For a fixed invariant mass squared $s$ of the $\omega\pi$ system, the maximum value of the dilepton invariant mass squared is therefore determined by $q_{\rm max}^2 = (m_B-\sqrt{s})^2$. To obtain the double differential decay width with respect to the hadronic invariant mass squared $s$ and the dilepton invariant mass squared $q^2$,  the angular variables appearing in Eq.~\eqref{sitidT} need to be integrated over. Upon integration over the full angular phase space, only terms with nonvanishing angular integrals survive, while the remaining interference terms vanish. Thus, the double differential decay width is obtained as:
\begin{align}
&\frac{d\Gamma(B\to b_1(1235)(\to \omega\pi)\, \ell^+ \nu\bigr)}{ds dq^2}
 = \frac{2}{9}\frac{G_F^2 |V_{ub}|^2 p_{\omega\pi}}{1024\pi^5 m_{B}^2\, \beta(s)}
\nonumber\\
&\,\,\,\,\,\,\,\,\,\,\,\times \big[|\mathcal{F}_{11}(s,q^2)|^2 + |\mathcal{F}_{21}(s,q^2)|^2 + |\mathcal{F}_{31}(s,q^2)|^2
\big],
\label{eq:sitiwidth}
\end{align}
Here, $\mathcal{F}_{11}(s,q^2)$, $\mathcal{F}_{21}(s,q^2)$, and $\mathcal{F}_{31}(s,q^2)$ denote the corresponding P-wave amplitudes. In the quark model, the $b_1(1235)$ resonance is classified as a $^1P_1$ state, corresponding to a $P$-wave axial-vector state~\cite{ParticleDataGroup:2026mpi, Godfrey:1985xj}. Therefore, only the $P$-wave contribution is retained. According to previous studies, $P$-wave amplitudes entering the angular distribution can be directly related to the helicity amplitudes describing the weak $B\to b_1(1235)$ transition ~\cite{Lee:1992ih, Richman:1995wm}. Therefore, the partial-wave amplitudes $\mathcal{F}_{i1}(s,q^2)$ ($i = 1,2,3$) can be expressed in terms of the helicity amplitudes as follows:
\begin{align}
\mathcal{F}_{11}(s,q^2) & =  2\sqrt{2}\alpha q H_0(s,q^2) {\cal A}(s),
\nonumber\\
\mathcal{F}_{21}(s,q^2)  & =  2\alpha q (H_+(s,q^2) + H_-(s,q^2)) {\cal A}(s),
\nonumber\\
\mathcal{F}_{31}(s,q^2)  & =  2\alpha q (H_+(s,q^2) - H_-(s,q^2)) {\cal A}(s).
\label{eq:ffs}
\end{align}
Here, $\alpha  =  [3 \pi \mathcal{B}_\omega / (p_0^* \Gamma^0_{b_1(1235)})]^{1/2}$. We take the total decay width $\Gamma^0_{b_1(1235)} = 0.142~\rm GeV$ and the branching fraction $\mathcal{B}_\omega  = \mathcal{B}(b_1(1235)\to \omega\pi) = 1$~\cite{BESIII:2024pwp}. The quantity $p_0^*$ denotes the magnitude of the three-momentum of the $\omega$ meson in the $b_1(1235)$ resonance rest frame, evaluated at the resonance pole mass. In addition, the line shape of the $b_1(1235)$ resonance can be described by a Breit-Wigner parametrization with a mass-dependent width~\cite{BaBar:2010vmf}. The corresponding line-shape amplitude is given by~\cite{BaBar:2010vmf,BESIII:2015hty}:
\begin{equation}
{\cal A}(s)  =  \frac{m_{b_1(1235)} \Gamma^0_{b_1(1235)}}{m_{b_1(1235)}^2-s-i m_{b_1(1235)} \Gamma_{b_1(1235)}(s)} \frac{B(p^*)}{B(p^*_0)}.
\end{equation}
The $\Gamma_{b_1(1235)}(s)$ denote the mass dependence width have the following form
\begin{equation}
\Gamma_{b_1(1235)} (s)  = \Gamma^0_{b_1(1235)} \Big(\frac{p^*}{p^*_0}\Big)^3 \frac{m_{b_1(1235)}}{\sqrt{s}}  \bigg[\frac{B(p^*)}{B(p^*_0)}  \bigg]^2
\end{equation}
The $B(p) = 1/(1+r^2_{\rm BW} p^{*2}_0)^{1/2}$ is the Blatt-Weisskopf damping factor. The three helicity amplitudes can then be related to the axial-vector and vector TFFs $V_{1,2}(q^2)$ and $A(q^2)$,
\begin{align}
\label{eq:sitiHi}
&H_{\pm}(s,q^2)  =  (m_B +\sqrt{s})V_1(q^2) \mp \frac{2 m_B p_{\omega\pi}}{m_B + \sqrt{s}} A(q^2),
\nonumber\\
&H_{0}(s,q^2) =  \frac{1}{2\sqrt{s} \sqrt{q^2}} \big[(m_{B}^2 - s - q^2)(m_B + \sqrt{s}) V_1(q^2)
\nonumber\\
&\qquad\qquad - \frac{4 m_{B}^2 p_{\omega\pi}^2}{m_B + \sqrt{s}} V_2(q^2) \big],
\end{align}
where the TFFs are evaluated at the pole mass of the $b_1(1235)$ resonance and have no explicit dependence on $s$. We find that $V_2(q^2)$ contributes only to $H_0(q^2)$, whereas $A(q^2)$ contributes only to $H_{\pm}(q^2)$. By contrast, $V_1(q^2)$ contributes to all three helicity amplitudes, $H_{\pm}(q^2)$ and $H_0(q^2)$. At large values of $q^2$, the contribution from $V_1(q^2)$ becomes increasingly important.

Based on the above theoretical analysis, the three TFFs serve as the main nonperturbative inputs for the calculation. A precise understanding of the behavior of these TFFs is essential for two reasons: it enables accurate predictions of the decay branching fraction and provides a theoretical basis for extracting the CKM matrix element $|V_{ub}|$ from experimental measurements. The TFFs are indispensable for calculating the above observable physical quantities. Under the LCSR framework, the TFFs are derived from the vacuum-to-meson correlation function, and their dominant form is given as:
\begin{align}
\hspace{-0.1cm}\Pi_\mu(p, q)  =  i \int d^4x e^{iq\cdot x} \langle b_1(1235)(p, \lambda) | T\{ j_\mu (x), j_B^\dagger(0) \} |0\rangle,
\label{Ep.Guanlian}
\end{align}
where $j_\mu (x)  =  \bar{u}(x) \gamma_\mu(1 - \gamma_5)b(x)$ and $j_B^\dagger(x) =i\bar{b}(x)(1+\gamma _5) q_2(x)$. This choice removes contributions from chiral-odd LCDAs, so that the non-perturbative inputs in the TFFs are mainly determined by the chiral-even LCDAs. Following the standard QCDSR, we insert a complete set of states with the same quantum numbers as the $B$ meson into the correlation function defined in Eq.~\eqref{Ep.Guanlian}. In this step, we use the matrix element $\langle B (p+q)|\bar{b} i \gamma _5 q_2|0 \rangle  =  m_{B}^{2}f_{B}/m_b$, where $f_B$ is the $B$-meson decay constant, together with the Lorentz-invariant TFFs. After isolating the pole term of the lowest pseudoscalar $B$-meson and using a dispersion relation to model the contributions from higher resonances and the continuum, we obtain the hadronic representation of the correlation function. In the spacelike $q^2$-region, the correlation function defined in Eq.~\eqref{Ep.Guanlian} can be calculated using QCD. Specifically, we perform an operator product expansion (OPE) near the light cone, where $x^2\sim 0$. This expansion is valid when $(p+q)^2 - m_b^2 \ll 0$ and the momentum transfer satisfies $q^2 \sim \mathcal{O}(1~{\rm GeV}^2 ) \ll m_b^2$. We then apply a Borel transformation with respect to $(p+q)^2$ to enhance the ground-state contribution relative to higher resonances. The continuum contribution is treated using quark-hadron duality, allowing the OPE representation to be matched to the hadronic representation. Within the LCSR framework, this procedure gives the following general expressions for the TFFs:
\begin{align}
{\cal R}_i(q^2)&  =  {\cal N}_i \! \bigg[ A_i \phi_{2;{b_1(1235)}}^\bot(u,\mu)  \! +  \! B_i\psi_{3;{b_1(1235)}}^\|(u)  \! -  \!  C_i\phi_{4;{b_1(1235)}}^\bot(u)
\nonumber\\
& - D_i I_L(u) - E_i H_3(u)  + G_i \Phi_{3;{b_1(1235)} }^\bot (\underline \alpha) + J_i \Psi_{4;{b_1(1235)}}^\bot (\underline \alpha)
\nonumber\\
&+ Q_i \widetilde \Psi_{4;{b_1(1235)}}^ \bot (\underline \alpha)\bigg], \label{Eq:ViTFFs}
\end{align}
where $i = 1, 2, 3$ corresponding to $V_1(q^2)$, $V_2(q^2)$ and $A(q^2)$ in the axial-vector and vector TFFs. For $i = 0$, $V_0(q^2)$ in the axial-vector TFFs satisfies ${\cal R}_0(q^2) = V_3(q^2)-V_0(q^2)$, where $V_3(q^2)$ is an auxiliary TFF. The explicit expressions for all parameters appearing in the analytical formulas for the TFFs are presented in Appendix~\ref{appendixA}.

For the two-particle twist-3 LCDAs $\psi _{3;b_1(1235)}^{\parallel}(x,\mu )$ and $\phi _{3;b_1(1235)}^{\parallel}(x,\mu)$, the analysis is more complicated than that for pseudoscalar states. This is because axial-vector states have different polarization states and their mass cannot be ignored. As a result, handling these functions becomes more difficult. In 2007, the twist-3 LCDAs of axial-vector states were systematically studied~\cite{Yang:2007zt}. Using a conformal partial-wave expansion and the QCD equations of motion, which showed that $\psi _{3;b_1(1235)}^{\parallel}(x,\mu )$ and $\phi _{3;b_1(1235)}^{\parallel}(x,\mu)$ are not independent. Instead, they can be expressed in terms of simpler quantities: the Wandzura-Wilczek (WW) part derived from the twist-2 LCDA, the three-particle LCDAs, and the terms from quark mass corrections. Among them, the contribution from the three-particle LCDAs is very small, we can safely ignore it. The quark mass correction terms involve a factor $\widetilde{\delta }_{\pm}$, which depends on the quark mass inside the meson. For the $b_1(1235)$ resonance, this factor is basically zero~\cite{Ball:1998sk}. As a result, the WW approximation is adopted in this paper. Under this approximation, the two-particle twist-3 LCDAs $\psi _{3;b_1(1235)}^{\parallel}(x,\mu )$ and $\phi _{3;b_1(1235)}^{\parallel}(x,\mu)$ can be approximately expressed as convolution integrals of the transverse twist-2 LCDA $\phi_{2;b_1(1235)}^\bot(x,\mu)$. The explicit expressions are given in Ref.~\cite{Yang:2007zt}. Therefore, the twist-3 two-particle LCDAs  required in the subsequent calculations can all be determined from $\phi_{2;b_1(1235)}^\bot(x,\mu)$  under the WW approximation. Using the higher-twist LCDAs given in Ref.~\cite{Ball:1998fj} within the simplest self-consistent approximation satisfying the QCD equations of motion, we obtain the expressions for $\Phi_{3;b_1(1235)}^{\bot}(\alpha_i)$, $\psi_{4;b_1(1235)}^{\bot}(x)$ and $\phi_{4;b_1(1235)}^{\bot}(x)$. The values of the coefficients entering these expressions are given in Ref.~\cite{Ball:2007rt}.

The correlation functions in Eq.~\eqref{Eq:ViTFFs}, together with the WW approximation, indicate that the dominant contributions to the TFFs arise from the transverse twist-2 LCDA of the $b_1(1235)$ resonance. We therefore employ the light-cone harmonic oscillator (LCHO) model to construct this LCDA. The twist-2 LCDA $\phi_{2;b_1(1235)}^{\bot}(x,\mu_0)$ can be obtained from the corresponding light-cone wave function (LCWF) $\psi_{2;b_1(1235)}^{\bot}(x,\mathbf{k}_\bot)$ by integrating over the transverse momentum. The relation between the LCDA and LCWF is given by:
\begin{align}\label{a20}
\hspace{-0.2cm}\phi_{2;b_1(1235)}^{\bot}(x, \mu) =   \frac{2\sqrt{3}}{\tilde{f}_{b_1(1235)}^{\bot}}\int_{|\mathbf{k}_{\perp}|^2 \leq \mu_0^2}\!\frac{d\mathbf{k}_{\bot}}{16\pi^3} \psi_{2;b_1(1235)}^\bot(x, \mathbf{k}_\bot),
\end{align}
where $\tilde{f}_{b_1(1235)}^\bot = f_{b_1(1235)}^\bot/\sqrt{3}$ and $k_\bot$ represents the transverse momentum of the $b_1(1235)$ resonance. Following the the Brodsky-Huang-Lepage (BHL) description~\cite{Zhong:2011rg,Wu:2007rt,Huang:2004su}, the light meson wave function $\psi _{2;b_1(1235)}^\bot(x,k_\bot)$ can be broken down into a spin wave function $\chi _{2;b_1(1235)}(x,k_\bot)$ and a spatial wave function $\Psi _{2;b_1(1235)}^{\rm R}(x,k_\bot)$, which can be written as
\begin{equation}
\psi_{2;b_1(1235)}^\bot(x, \mathbf{k}_\bot)  = \sum_{h_1 h_2} \chi_{b_1(1235)}^{h_1 h_2}(x, \mathbf{k}_\bot) \Psi_{2;b_1(1235)}^{\rm R} (x, \mathbf{k}_\bot),
\end{equation}
In this expression, the correct relativistic spin structure is provided by $\chi_{b_1(1235)}^{h_1 h_2}(x, \mathbf{k}_\bot)$, and the longitudinal and transverse momentum distributions of the quark-antiquark pair are described by $\Psi _{2;b_1(1235)}^{\rm R}(x,k_\bot)$. The symbols $h_1$ and $h_2$ refer to the helicities of the quark and antiquark, respectively. The function $\chi_{b_1(1235)}^{h_1 h_2}(x, \mathbf{k}_\bot)$ describes the coupling structure of the quark and antiquark spins, and it is determined by the Wigner Melosh rotation~\cite{Huang:1994dy,Huang:2004su,Wu:2005kq}. The spin-orbit coupling information is incorporated into the corresponding projection factors and wave-function definitions through the Wigner-Melosh rotation. The spin wave function then takes the simplified form $\chi_{b_1(1235)}^{h_1 h_2}(\mathbf{k}_{\perp}) = \hat m_q/(\mathbf{k}_{\perp}^2 + \hat m_q^2)^{1/2}$, where $\hat m_q =  300 ~ {\rm MeV}$ is the constituent quark mass. Brodsky-Huang-Lepage (BHL) established a scheme by comparing the off-shell propagators in the rest frame and the light-front representation. In this scheme, the harmonic oscillator wave function in the rest frame is mapped to the light-front momentum space~\cite{Brodsky:1980vj}. Therefore, the spatial wave function $\Psi _{2;b_1(1235)}^{R}(x,k_\bot)$ can be split into two parts. The first depends only on the longitudinal momentum fraction $x$ and is written as $\varphi(x)$. The second depends on $k_\bot$ and comes from the harmonic oscillator solution that describes the meson in its rest frame $\Psi_{2;b_1(1235)}^{\rm R} (x, \mathbf{k}_\bot)  =  A_{2;b_1(1235)}^\bot \varphi(x) \exp [-b_{2;b_1(1235)}^\bot (\mathbf{k}_\bot^2 + \hat m_q^2) / (x\bar{x})]$ with $\bar x  =  (1-x)$, where $A_{2;b_1(1235)}^\bot$ is the normalization constant, $b_{2;b_1(1235)}^\bot$ is the harmonic parameter. When quark-spin effects are taken into account, the Lorentz boost induces an additional longitudinal correction to the LCDA, which is incorporated through the function $\varphi(x)$. This function can be expanded in terms of Gegenbauer polynomials as $\varphi(x)  =  1 + B_{2;b_1(1235)}^\bot C_2^{3/2}(\xi)$ with $\xi =  (2x-1)$, where $C_2^{3/2}(\xi)$ is a Gegenbauer polynomial and $B_{2;b_1(1235)}^\bot$ plays an important role in determining the longitudinal part of the distribution. Finally, by integrating over the transverse momentum using Eq.~\eqref{a20}, we can determine the twist-2 LCDA $\phi_{2;b_1(1235)}^{\perp}(x, \mu)$ within the LCHO model. The result is
\begin{align}
&\hspace{-0.2cm}\phi_{2;b_1(1235)}^{\perp}(x, \mu)  =  \frac{A_{2;b_1(1235)}^{\perp} \sqrt{3x\bar{x}\hat m_q}}{8\pi^{3/2} f_{b_1(1235)}^{\perp} b_{2;b_1(1235)}^{\perp}} \left[ 1 \! + \!  B_{2;b_1(1235)}^{\perp} C_2^{3/2}(\xi) \right]
\nonumber\\
&\times \!\bigg[ {\rm Erf} \!  \bigg( b_{2;b_1(1235)}^{\perp}  \! \sqrt{\frac{\mu^2 \!  +  \! \hat m_q^2}{x\bar{x}}} \bigg)  \! - \!  {\rm Erf} \!  \bigg( b_{2;b_1(1235)}^{\perp} \!  \sqrt{\frac{\hat m_q^2}{x\bar{x}}} \bigg)  \! \bigg],\label{FB2}
\end{align}
where $\operatorname{Erf}(x)  =  2\int_0^x{e^{-t^2}dt}/\sqrt{\pi}$ denotes the error function, it is obtained by integrating over the transverse momentum and rewriting the integral in Gaussian form. The integral result can be expressed as the difference between two error functions, with the first evaluated at the upper limit of the transverse-momentum integral $|\mathbf{k}_\perp| = \mu$, and the second at the lower limit $|\mathbf{k}_\perp| = 0$. The difference between these two error functions represents the contribution to the twist-2 LCDA from transverse momentum below the factorization scale $\mu$. This form in Eq.~(\ref{FB2}) has two useful features. First, it approaches the asymptotic form $\phi _{2;b_1(1235)}^\bot(x,\infty)  = 6x\bar{x}$ in the limit of large $\mu$. Second, it is exponentially suppressed near the endpoints, thereby suppressing endpoint singularities in convolution integrals. In addition, three model parameters remain to be determined $A_{2;b_1(1235)}^\bot$, $b_{2;b_1(1235)}^\bot$, and $B_{2;b_1(1235)}^\bot$. To do that, we use some extra conditions: (i) The normalization condition of the twist-2 LCDA for the $b_1(1235)$ resonance; (ii) The average value of the transverse momentum squared $\langle \mathbf{k}_{\perp}^2 \rangle_{2;b_1(1235)}^{1/2} = 0.37~{\rm GeV}$~\cite{Guo:1991eb}; (iii) The first Gegenbauer moments of the twist-2 LCDA. For this purpose, we adopt the value $a_{2;b_1(1235)}^\bot(\mu _0)  = 0.03\pm 0.19$ at the initial scale $\mu _0 = 1~{\rm GeV}$~\cite{Yang:2007zt}. The above theoretical framework lays the foundation for calculating the physical observables considered in this work.

\section{Numerical Analysis}\label{sec:analysis}
\begin{table}[t]
\begin{center}
\renewcommand{\arraystretch}{1.2}
\footnotesize
\caption{The three model parameters of $b_1(1235)$ resonance transverse twist-2 LCDA, {\it i.e.} $A_{2;b_1(1235)}^\bot$, $B_{2;b_1(1235)}^\bot$ and $b_{2;b_1(1235)}^\bot$ at initial and typical scale $\mu_{0,k}$. From which the upper limits, central values, lower limits are present, respectively.}
\setlength{\tabcolsep}{5pt}
\label{table:modelparameter}
\begin{tabular}{l @{\hspace{1cm}} l  @{\hspace{1cm}} l  @{\hspace{1cm}} l}
\hline
 &$A_{2;b_1(1235)}^\bot $&$B_{2;b_1(1235)}^\bot $ &$b_{2;b_1(1235)}^\bot $  \\
\hline
                     &+25.877                    &+0.182                    &$-0.574$      \\
$\mu_0$ ~~~          &+29.671                   &+0.012                    &$-0.626$   \\
                     &+32.401                   &$-0.175$                    &$-0.662$    \\
\hline
                     &+24.703                    &+0.152                    &$-0.584$  \\
$\mu_k$  ~~~         &+27.604                    &+0.022                    &$-0.623$   \\
                     &+29.910                   &$-0.118$                    &+0.653    \\
\hline
\end{tabular}
\end{center}
\end{table}

\begin{figure}[t]
\begin{center}
\includegraphics[width = 0.4\textwidth]{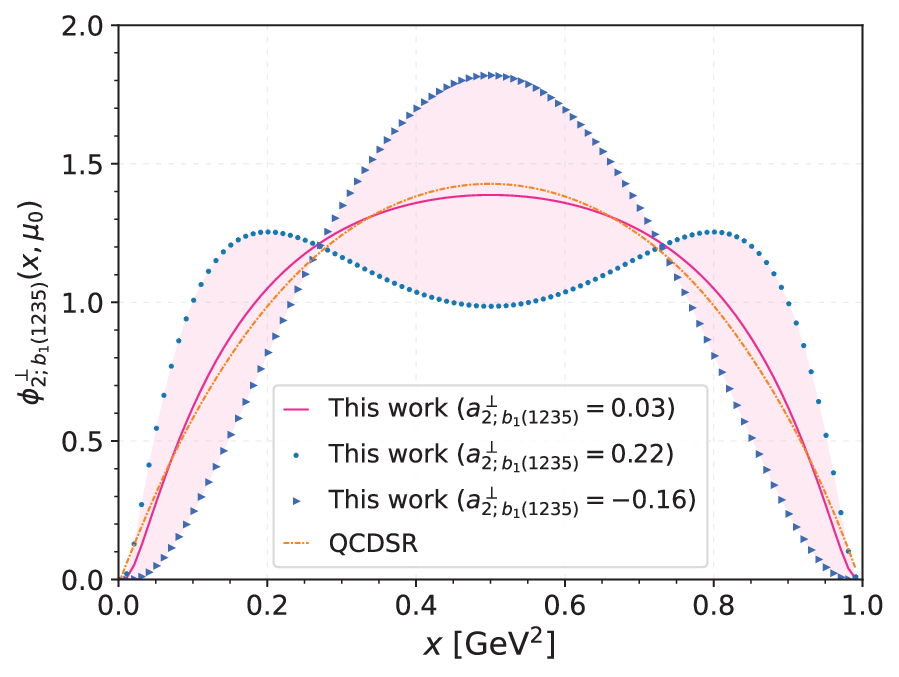}
\caption{The behavior of $\phi _{2;b_1(1235)}^\bot\left(x,\mu _0 \right)$ in LCHO model within uncertainties.The dashed line represents the previous QCDSR~\cite{Yang:2007zt} prediction.}
\label{fig:DAb1}
\end{center}
\end{figure}
\begin{table*}[t]
\footnotesize
\begin{center}
\caption{The $B\to b_1(1235)$ TFFs $V_1(0), V_2(0)$ and $A(0)$ at the large recoil . The predictions of other theories are used for comparison.}
\label{table:A1A2V0}
\begin{tabular}{l @{\hspace{1.5cm}} l @{\hspace{1.5cm}} l @{\hspace{1.5cm}} l @{\hspace{1.5cm}} l}
\hline
& $V_1(0)$ & $V_2(0)$ & $V_0(0)$ & $A(0)$\\
\hline
This work & $+0.338^{+0.055}_{-0.052}$    & $+0.187^{+0.035}_{-0.034}$    & $+0.563^{+0.082}_{-0.070}$        & $+0.517^{+0.079}_{-0.084}$   \\
pQCD~\cite{Li:2009tx}& $+0.330^{+0.070+0.010}_{-0.060-0.020}(50)$    & $+0.030(10)(0)(20)$  & $+0.450^{+0.100}_{-0.090}(10)(40)$   & $+0.190(40)(10)(30)$   \\
LCSR(I)~\cite{Zeng:2025gft}     &  $-0.096^{+0.014}_{-0.017}$    &  $+0.166^{+0.024}_{-0.023}$      &  $-0.597^{+0.065}_{-0.070}$   &  $-0.189^{+0.055}_{-0.068}$ \\
LCSR(II)~\cite{Sun:2011ssd}     &  $-0.290(90)(60)$    &  $-0.170(60)(40)$      &  $-0.050(50)(50)$   &  $-0.17(6)(4)$ \\
LCSR(III)~\cite{Yang:2008xw}     &  $-0.200(40)$    &  $-0.090(20)$      &  $-0.390(70)$   &  $-0.250(50)$ \\
LFQM~\cite{Cheng:2003sm}     &  $+0.180$    &  $-0.030$    &  $+0.390$    &  $+0.100$ \\
\hline
\end{tabular}
\end{center}
\end{table*}
Before performing the numerical analysis, we adopt the following input parameters from PDG\cite{ParticleDataGroup:2026mpi}. The masses of charged and neutral $B$-mesons and $b_1(1235)$-resonance are taken as $m_{B^+} = 5279.42\pm 0.08 \ \mathrm{MeV}$, $m_{B^0} = 5279.72\pm 0.08 \ \mathrm{MeV}$ and $m_{b_1(1235)} = 1229.5\pm 3.2 \ \mathrm{MeV}$, respectively. For the quark masses, we use $m_b = 4.18_{-0.03}^{+0.04} \ \mathrm{GeV}$, $m_d = 4.76\pm 0.07 \ \mathrm{MeV}$ and $m_u = 2.16\pm 0.07 \ \mathrm{MeV}$. The masses of the $\omega$ and $\pi$ mesons are taken as $m_{\omega} = 782.66\pm0.13 \ \mathrm{MeV}$ and $m_{\pi} = 139.57\pm0.00018 \ \mathrm{MeV}$. The decay constant $f_{B^+} = f_{B^0} = 0.160\pm 0.019{\rm GeV}$ \cite{Fu:2014pba}, and $f_{b_1(1235)}^\bot = 0.180\pm 0.0008 \ \mathrm{GeV}$ \cite{Yang:2007zt}.
For $B\to b_1(1235)(\to \omega\pi)$ transition, we take the typical process scale $\mu_k = (m_{B}^2-m_{b}^2)^{1/2}\simeq 3.2 \ \mathrm{GeV}$. The transverse twist-2 LCDA $\phi_{2;b_1(1235)}^\bot(x,\mu)$ plays a key role in determining the TFFs, both directly and indirectly. We therefore investigate the behavior of the transverse twist-2 LCDA $\phi _{2;b_1(1235)}^\bot(x,\mu)$ firstly. Using the three conditions mentioned before: normalization, average transverse momentum squared, and Gegenbauer moment, we can fix the unknown parameters for $\phi_{2;b_1(1235)}^\bot(x,\mu)$. To obtain the LCDA parameters at any scale, the Gegenbauer moment $a_{n}^{2;b_1(1235)} (\mu)$ can be evolved from initial scale $\mu_0$ to typical process scale $\mu_k$ by using the renormalization group equation
\begin{align}
a_{n}^{2;b_1(1235)} (\mu)  =  a_{n}^{2;b_1(1235)} (\mu_0)\bigg [ \frac{\alpha_s(\mu_0)}{\alpha_s(\mu_k)}\bigg]^{-\frac{\gamma_n }{2\beta _0}}.
\end{align}
The one-loop anomalous dimensions are given by~\cite{Ball:2004ye, Braun:2003rp}:
\begin{align}
\gamma_n & =  C_F \bigg( 1 - \frac{2}{(n+1)(n+2)} + 4 \sum_{j = 2}^{n+1} \frac{1}{j} \bigg).
\end{align}
Here, $\beta_0  =  11 - 2/3 n_f$, where $n_f$ denotes the number of active quark flavors. The resulting numerical values are listed Table~\ref{table:modelparameter}. The resulting behavior of $\phi _{2;b_1(1235)}^\bot(x,\mu )$ is shown in Fig.~\ref{fig:DAb1}, where the uncertainty comes from the error in $a_{2;b_1(1235)}^\bot( \mu_0 )$. The transverse twist-2 LCDA $\phi _{2;b_1(1235)}^\bot(x,\mu _0 )$ obtained in this work is presented in Fig.~\ref{fig:DAb1}, with the solid curve representing the central value and the shaded band indicating the corresponding uncertainty. For comparison, the QCDSR prediction based on the truncated Gegenbauer expansion is also included~\cite{Yang:2007zt}. Our result is in good agreement with the QCDSR~\cite{Yang:2007zt} prediction, with both exhibiting a single-peak structure symmetric about $x = 0.5$, and agreement with the theoretical expectation. Our numerical result is slightly higher than the QCDSR~\cite{Yang:2007zt} prediction, but it is still within the error range. This discrepancy may arise from differences in model construction, the choice of input parameters, and uncertainty propagation. Overall, the results presented in Fig.~\ref{fig:DAb1} support the applicability of the LCHO model in constructing the transverse twist-2 LCDA of the $b_1(1235)$ resonance and provide a useful theoretical input for the subsequent calculation of the TFFs. In addition, the calculation of the TFFs for the $B\to b_1(1235)$ transition requires two important input parameters: the continuum threshold $s_0$ and the Borel parameter $M^2$. In the QCD sum rule approach~\cite{Zhong:2023cyc,Tian:2024ubt,Wang:2025oix}, we use the following criteria to determine them:
\begin{itemize}
    \item The contribution from the continuum is less than $35\%$.
    \item The contribution from higher-twist LCDAs does not exceed $5\%$.
    \item The value of the TFFs remains stable within the Borel window.
\end{itemize}
Based on the constraints mentioned above, we choose the continuum thresholds as $s^{V_{0,1,2}}_0 =  30.0(1)~{\rm GeV}^2$ and $s^{A}_0  =  32.0(1)~{\rm GeV}^2$. The corresponding Borel parameters are chosen as
$M^2_{V_1} = 5.6(1)~{\rm GeV}^2, M^2_{V_2} = 11.0(1)~{\rm GeV}^2, M^2_{V_0}  =  7.8(1)~{\rm GeV}^2$, and $M^2_A  =  6.0(1)~{\rm GeV}^2$. Using the above parameters and Eq.~\eqref{Eq:ViTFFs}, we can determine the TFFs for $B\to b_1(1235)$ at large recoil. The results are presented in Table~\ref{table:A1A2V0} and compared with the theoretical predictions from pQCD~\cite{Li:2009tx}, LCSR(I)~\cite{Zeng:2025gft}, and LCSR(II)~\cite{Sun:2011ssd}. The uncertainties in our predicted TFFs are controlled within the range of $10\% - 15\%$. Our results for $V_1(0)$ agrees well with the pQCD prediction~\cite{Li:2009tx}, while our result for $V_2(0)$ is close to the LCSR(I) prediction~\cite{Zeng:2025gft}. For the remaining TFFs, our predictions have some differences compared to LCSR(I)~\cite{Zeng:2025gft} and LCSR(II)~\cite{Sun:2011ssd}. These discrepancies may arise from differences in the choices of chiral currents and the input distribution amplitudes adopted in the respective approaches.
\begin{table}[t]
\renewcommand{\arraystretch}{1}
\footnotesize
\begin{center}
\caption{The ratio $r_V$ and $r_2$ of $B\to b_1(1235)$ TFFs. For comparison, other theoretical results are also included.}
\label{table:ratio}
\begin{tabular}{l@{\hspace{1em}} l@{\hspace{1em}} l}
\hline
& $r_V$~~~~~& $r_2$\\
\hline
This work& $+1.530_{-0.016}^{+0.013}$    & $+0.553_{-0.015}^{+0.012}$ \\
pQCD~\cite{Li:2009tx}& $+0.576_{-0.020-0.001-0.005}^{+0.005+0.012+0.003}$    & $+0.091_{-0.016-0.058}^{+0.010+0.057}$     \\
LCSR(I)~\cite{Zeng:2025gft}     &  $+1.969_{-0.335}^{+0.305}$    &  $-1.729_{-0.588}^{+0.464}$\\
LCSR(II)~\cite{Sun:2011ssd}     &  $+0.586_{-0.036-0.021}^{+0.019+0.014}$    &  $+0.586_{-0.036-0.021}^{+0.019+0.014}$ \\
LCSR(III)~\cite{Yang:2008xw}    &  $+1.250_{-0.000}^{+0.000}$    &  $+0.450_{-0.012}^{+0.008}$ \\
LFQM~\cite{Cheng:2003sm}     &  $+0.556$    &  $-0.167$ \\
\hline
\end{tabular}
\end{center}
\end{table}
\begin{figure}[b]
\begin{center}
\includegraphics[width = 0.43\textwidth]{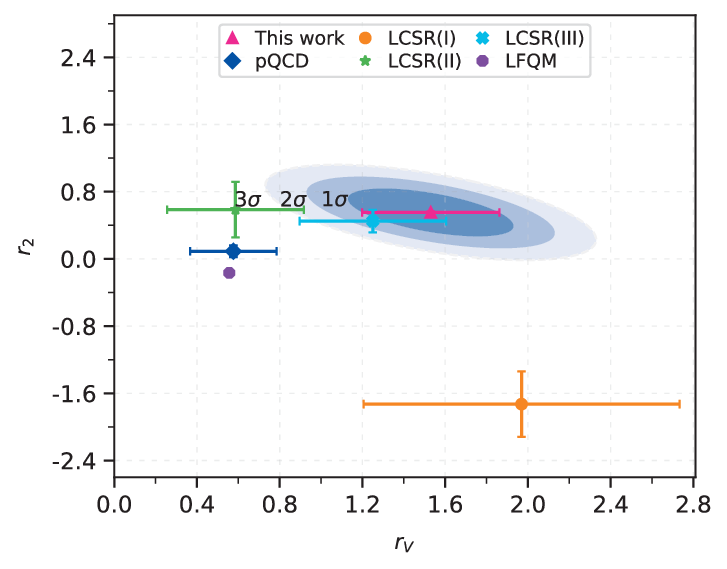}
\caption{The $1\sigma$ confidence ellipses for the TFFs ratios $r_V$ and $r_2$ in the decay $B\to b_1(1235)$ are shown. The empirical value $rho=-0.57$ is adopted from Ref.~\cite{BaBar:2010vmf} to determine the tilt of the ellipses. We also compare our results with predictions from other theoretical approaches.}
\label{fig:RVR2}
\end{center}
\end{figure}
\begin{figure*}[t]
\begin{center}
\includegraphics[width = 0.84\textwidth]{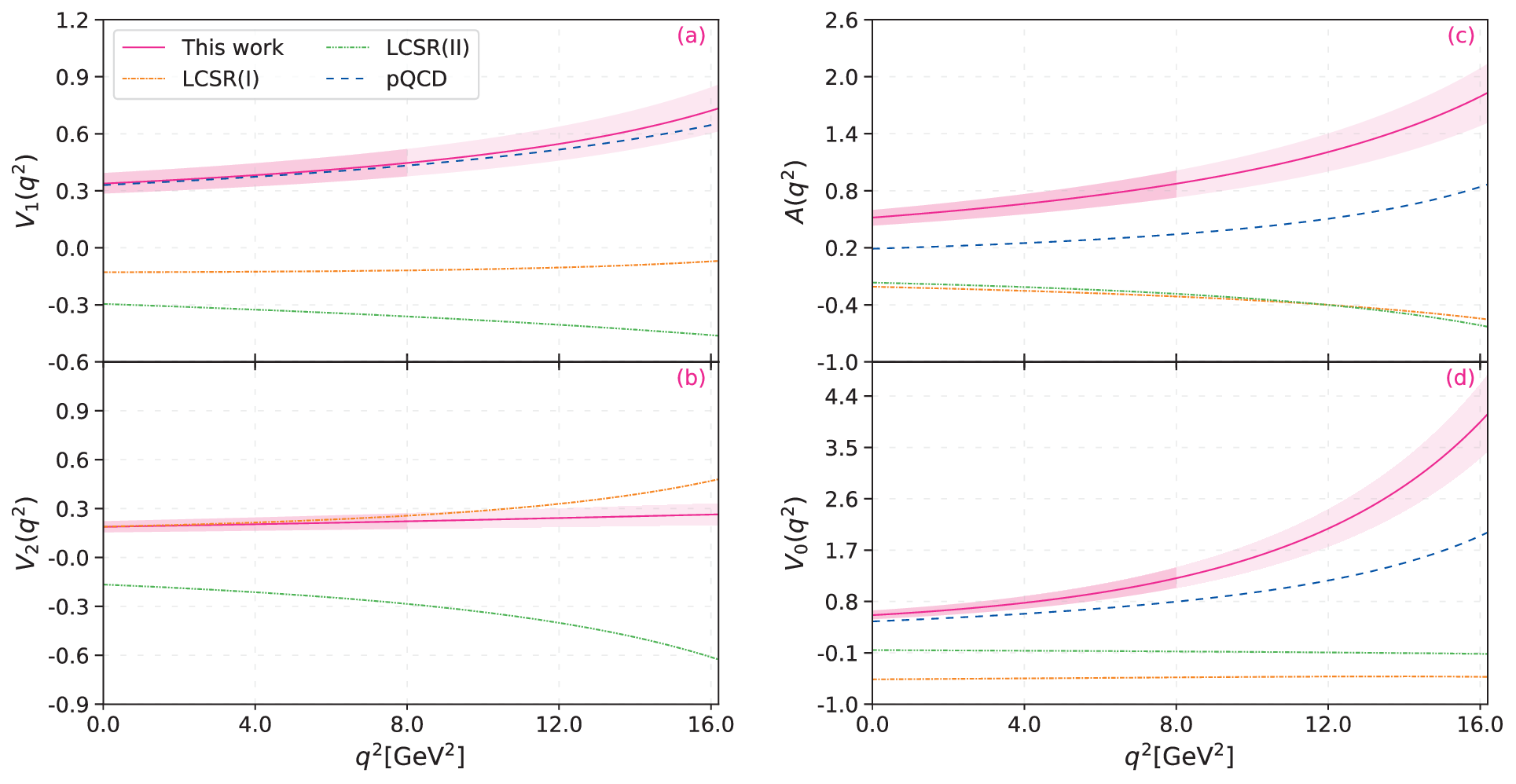}
\caption{The extrapolated TFFs $V_{0,1,2}(q^2)$ and $A(q^2)$ versus $q^2$, where the solid line represents the central value and the shaded bands represent the corresponding uncertainties. In addition, the results are also compared with the predictions from LCSR(I)~\cite{Zeng:2025gft}, LCSR(II)~\cite{Sun:2011ssd} and pQCD~\cite{Li:2009tx} in the figure.}
\label{fig:b1TFFs}
\end{center}
\end{figure*}
Additionally, we can define two particularly interesting ratios in the large-recoil region, $r_V = A(0)/V_1(0)$ and $r_2 = V_2(0)/V_1(0)$. These ratios can be obtained without relying on the total decay width or the CKM matrix elements, and they effectively remove common uncertainties from the overall normalization of the TFFs. This makes them more sensitive to differences in how various models describe the internal dynamics of hadrons. These ratios are listed in Table~\ref{table:ratio}, together with other theoretical predictions. Different theoretical approaches employ different methods and input parameters. In this work, we obtain the central values $r_V = 1.530_{-0.352}^{+0.332}$ and $r_2 = 0.553_{-0.135}^{+0.134}$. In particular, our prediction for $r_V$ differs appreciably from those of pQCD~\cite{Li:2009tx}, LCSR (II)~\cite{Sun:2011ssd}, and LFQM~\cite{Cheng:2003sm}. In general, these approaches predict smaller values of $r_V$ than our result. The predictions for $r_2$ also differ to varying degrees among different theoretical approaches, indicating that the relative magnitudes of the TFFs in the $B\to b_1(1235)$ transition remain somewhat model-dependent. To facilitate the comparison of theoretical predictions for $r_V$ and $r_2$, we present the corresponding confidence ellipse in Fig.~\ref{fig:RVR2}, which illustrates the joint probability distribution of the two ratios obtained in this work. The comparison reveals differences among theoretical predictions for the three TFFs at the large recoil point. To assess the reliability of these results, one can examine the overall behavior of the TFFs across the full kinematic range.

The LCSR method used to calculate the $B\to b_1(1235)$ TFFs is reliable mainly in the low and intermediate $q^2$ regions, but the decay width and branching fraction require knowledge of how the TFFs behave across the entire physical $q^2$. To analytically continue the TFFs to the full allowed physical region, we employ the simplified series expansion (SSE)~\cite{Bourrely:2008za}, given by  $F_{i}(q^2) = \frac{1}{1-q^2/m^2_{R,i}}\sum_{k = 0,1,2}\beta_{k,i} z^k(q^2,t_0)$. Where $\beta _{k,i}$ is a coefficients, the conformal variable $z(q^2, t_0)$ and the reference point $t_0$  follow the standard definitions given in Ref.\cite{Bourrely:2008za}. This parametrization provides a systematic and largely model-independent description of the semileptonic TFFs. It allows the near-threshold behavior of the form factors to constrain the expansion coefficients while preserving their analytic structure.
\begin{table}[t]
\renewcommand{\arraystretch}{1}
\footnotesize
\begin{center}
\caption{The fitted parameters $m_{R,i}~({\rm GeV})$, $\beta_{k,i}$ $(k = 0, 1, 2)$ and quality of extrapolation $\Delta$ for $B\to b_1(1235)$ TFFs $V_{0,1,2}(0)$ and $A(0)$. All LCSR parameters are set to be their central values.}
\setlength{\tabcolsep}{2pt}
\label{table:fitParameter}
\begin{tabular}{l@{\hspace{2em}} c@{\hspace{2em}} c@{\hspace{2em}} c@{\hspace{2em}} c}
\hline
  & $~V_1(0)$   & $~V_2(0)$   & $~A(0)$    & $~V_0(0)$\\
\hline
$m_{R,i}$   &$~5.726$     &~5.726   &~5.324   &~5.279\\
$\beta_{0,i}$   &~0.338     &~0.188  &~0.518   &~0.560\\
$\beta_{1,i}$   &0.460     &0.519  &$-1.981$   &~$-2.843$\\
$\beta_{2,i}$   &1.781    &~$-1.034$   &0.745   &~55.865\\
$\Delta$    &$~0.030\%$  &$~0.806\%$  &$~0.007\%$  &$~0.145\%$  \\
\hline
\end{tabular}
\end{center}
\end{table}

\begin{figure}[t]
\begin{center}
\includegraphics[width = 0.40\textwidth]{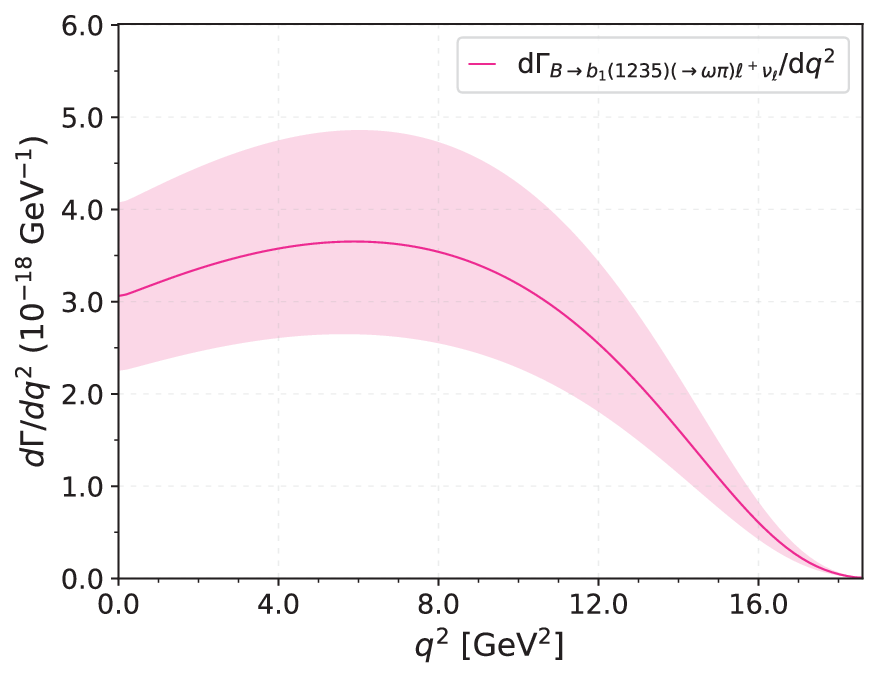}
\caption{The differential decay widths $d\Gamma /dq^2$ of $B \to b_1(1235)(\to \omega\pi)\ell^+ \nu_\ell$ are plotted as functions of $q^2$, the shaded area represents the uncertainty band of $d\Gamma(B \to b_1(1235)(\to \omega\pi)\ell^+ \nu_\ell)/dq^2$ for the corresponding channel.}
\label{fig:dGammadqb1omega}
\end{center}
\end{figure}

Additionally, the real coefficients $\beta _{k,i}$ are determined by requiring the quality of extrapolation $\Delta$ to be less than $1\%$. Using the input values reported in Ref.~\cite{Momeni:2020zrb} and by the PDG~\cite{ParticleDataGroup:2026mpi}, we obtain the fit parameters $m_{R,i}$ and the fit quality $\Delta$ for each form factor, as listed in Table~\ref{table:fitParameter}. In this case, all LCSR parameters are set to their central values. From the table, the values of $\Delta$ corresponding to the four TFFs are found to be $0.030\%$, $0.806\%$, $0.007\%$ and $0.145\%$, respectively, all of which are less than $1\%$. Among them, $A(0)$ has the best fit quality. Although the $\Delta$ value for $V_2(0)$ is relatively large, it is still significantly lower than $1\%$. This indicates that the adopted parametrization accurately reproduces the LCSR results for the TFFs within the fitting interval, providing a reliable basis for further investigating their behavior over the full kinematic region. Using the parametrization described above, we obtain the $q^2$ dependence of the TFFs over the full physical kinematic range, as shown in Fig.~\ref{fig:b1TFFs}. In Fig.~\ref{fig:b1TFFs}, the solid curves represent the central values, while the shaded bands indicate the corresponding uncertainties. We also compare our results with the predictions from LCSR (I)~\cite{Zeng:2025gft}, LCSR (II)~\cite{Sun:2011ssd}, and pQCD~\cite{Li:2009tx}. The different theoretical approaches exhibit noticeable differences in the extrapolation behavior of the TFFs. Our extrapolated TFFs show varying degrees of agreement with predictions from different theoretical approaches. In the low and intermediate $q^2$ regions, our result for $V_1(q^2)$ agrees well with the pQCD prediction~\cite{Li:2009tx}. Similarly, our result for $V_2(q^2)$ is close to the LCSR (I) prediction~\cite{Zeng:2025gft} in these regions, although the discrepancy gradually increases with $q^2$. In our predictions, both $V_0(q^2)$ and $A(q^2)$ remain positive and increase monotonically with $q^2$. This behavior differs markedly from the predictions of LCSR (I)~\cite{Zeng:2025gft} and LCSR (II)~\cite{Sun:2011ssd}. In contrast, the pQCD predictions~\cite{Li:2009tx} exhibit a similar $q^2$ dependence to that obtained in this work, although their overall magnitudes are smaller. These differences provide important clues for future experimental tests.
\begin{figure}[t]
\begin{center}
\includegraphics[width = 0.46\textwidth]{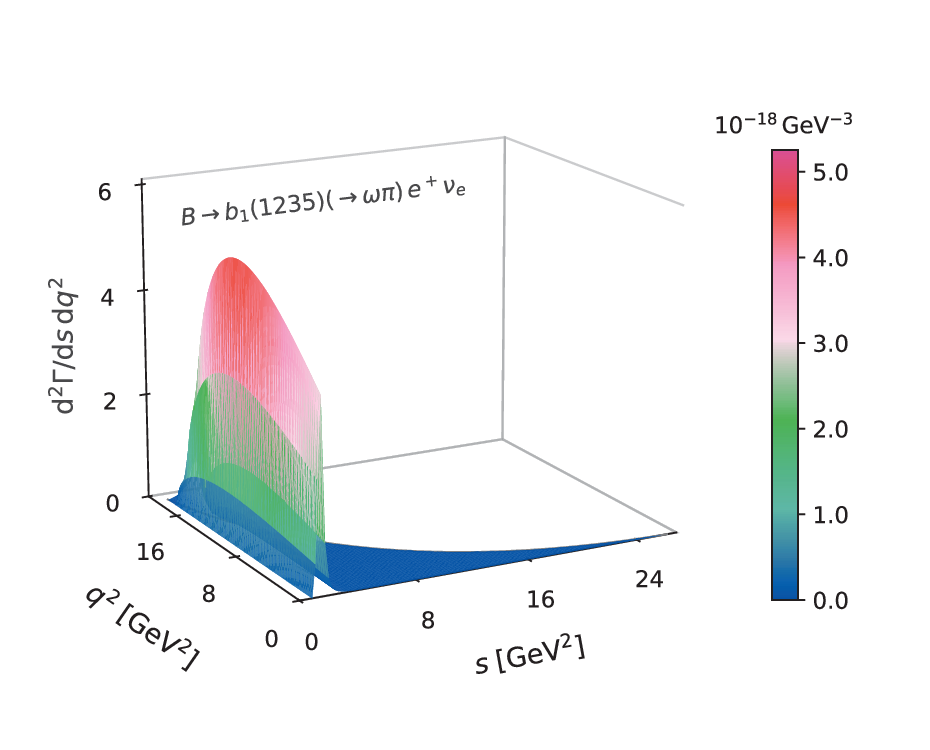}
\caption{For the process $B \to b_1(1235)(\to \omega\pi)e^+ \nu_e$, a three-dimensional distribution of the double differential width $d^2\Gamma/dsdq^2$ as a function of $s$ and $q^2$ is presented, where the height of the surface corresponds to the value of the differential decay width.}
\label{fig:dGammadqb1omega3D}
\end{center}
\end{figure}
\begin{table*}[t]
\begin{center}
\caption{Predictions for the branching ratios of the $B_{\ell4}$ decay $B\to b_1(1235)(\to \omega\pi)\ell^+\nu_\ell$ with $\ell = (e, \mu, \tau)$ and the three-body semileptonic decay $B\to b_1(1235)\ell^+\nu_\ell$ are given within the uncertainties (in unit: $10^{-4}$).}
\label{table:sitiBR}
\setlength{\tabcolsep}{3pt}
\footnotesize
\begin{tabular}{ l@{\hspace{1.2cm}} c@{\hspace{1.2cm}} c@{\hspace{1.2cm}} c@{\hspace{1.2cm}} c@{\hspace{1.2cm}} c }
\hline
 Decay channels  &This work & LCSR(I)~\cite{Zeng:2025gft} & LCSR(II)~\cite{Sun:2011ssd}  &pQCD~\cite{Li:2009tx} &CLFQM~\cite{Kang:2018jzg}\\
\hline
 $\mathcal{B}(B^+\to b_1^0(1235)(\to \omega\pi^0 )e^+\nu_e)$  &  $1.322^{+0.450}_{-0.370}$  & $-$    & $-$    & $-$    & $-$\\
  $\mathcal{B}(B^+\to b_1^0(1235)(\to \omega\pi^0 )\mu^+\nu_\mu)$  &  $1.321^{+0.451}_{-0.370}$  & $-$   & $-$  & $-$    & $-$\\
 $\mathcal{B}(B^+\to b_1^0(1235)(\to \omega\pi^0 )\tau^+\nu_\tau)$  &  $1.190^{+0.409}_{-0.338}$  & $-$    & $-$    & $-$    & $-$\\
 $\mathcal{B}(B^0\to b_1^-(1235)(\to \omega\pi^- )e^+\nu_e)$  &  $1.220^{+0.417}_{-0.342}$  & $-$    & $-$    & $-$\\
 $\mathcal{B}(B^0\to b_1^-(1235)(\to \omega\pi^- )\mu^+\nu_\mu)$  &  $1.219^{+0.416}_{-0.341}$  & $-$    & $-$    & $-$    & $-$\\
 $\mathcal{B}(B^0\to b_1^-(1235)(\to \omega\pi^- )\tau^+\nu_\tau)$  &  $1.099^{+0.377}_{-0.312}$  & $-$    & $-$    & $-$    & $-$\\
 \hline
 $\mathcal{B}(B^+\to b_1^0(1235) e^+\nu_e)$   &  $3.613^{+1.201}_{-1.045}$  & $2.353^{+0.597}_{-0.456}$    & $2.070^{+0.900}_{-0.730}$    & $-$    & $0.770\pm{0.170}$\\
 $\mathcal{B}(B^+\to b_1^0(1235)\mu^+\nu_\mu)$  &  $3.604^{+1.198}_{-1.043}$  & $2.339^{+0.587}_{-0.448}$     & $-$    & $-$    & $-$\\
 $\mathcal{B}(B^+\to b_1^0(1235) \tau^+\nu_\tau)$  &  $2.463^{+0.835}_{-0.705}$  & $-$    & $-$    & $-$    & $0.300\pm{0.070}$\\
 $\mathcal{B}(B^0\to b_1^-(1235) e^+\nu_e)$  &  $3.332^{+1.108}_{-0.940}$  & $-$    & $-$    & $-$    & $-$\\
 $\mathcal{B}(B^0\to b_1^-(1235) \mu^+\nu_\mu)$  &  $3.353^{+1.115}_{-0.965}$  & $2.166^{+0.544}_{-0.415}$    & $1.260^{+0.660}_{-0.540}$    & $-$\\
 $\mathcal{B}(B^0\to b_1^-(1235) \tau^+\nu_\tau)$  &  $2.248^{+0.762}_{-0.640}$  & $-$    & $-$    & $-$    & $-$\\
\hline
\end{tabular}
\end{center}
\end{table*}

The differential decay widths are calculated using these TFFs, with $| V_{ub} | = (3.82\pm 0.20)\times 10^{-3}$ adopted~\cite{ParticleDataGroup:2026mpi}. Detailed behavior is shown in Fig.~\ref{fig:dGammadqb1omega}. To calculate the total decay width of $B\to b_1(1235)(\to\omega\pi)\ell\nu_\ell$, the integration range over the invariant mass $s$ corresponding to Eq.~\eqref{eq:sitiwidth} is $(m_\omega+m_\pi)^2\leq s\leq (m_B-m_\ell)^2$. The maximum value of $q^2$ is given by $q_{\max}^2 = (m_{B}-(m_\omega+m_\pi))^2$. Overall, the peak in the differential decay-width distribution for $B \to b_1(1235)\ell^+\nu_{\ell}$ is relatively mild. This behavior can be attributed in part to the Breit-Wigner line shape of the $b_1(1235)$ resonance, which redistributes the available phase space and, through the kinematic correlations between $s$ and $q^2$, shifts the $q^2$ distribution toward lower values. In addition, a faster decline is observed in $B\to b_1(1235)(\to\omega\pi)\ell^+\nu_{\ell}$ near the endpoint, this is because the phase space of the $\omega\pi$ system is suppressed near the endpoint. To further illustrate the finite-width effects, the double-differential decay width $d^2\Gamma/(dsdq^2)$ is displayed in the $(s,q^2)$ plane in Fig.~\ref{fig:dGammadqb1omega3D}. The distribution is mainly concentrated near the $b_1(1235)$ resonance region, and the global maximum is found at $q^{2}\simeq10.1~\mathrm{GeV}^{2}$. As is seen in the figure, when $s$ increases from $1.08~\mathrm{GeV}^{2}$ to $1.98~\mathrm{GeV}^{2}$, the peak position of the $q^2$ distribution is shifted from about $11.40~\mathrm{GeV}^{2}$ to about $9.00~\mathrm{GeV}^{2}$, indicating that the shape of the $q^2$ distribution is changed with $s$. This behavior arises from the explicit dependence of the kinematic upper limit $q_{\max}^{2} = (m_B-\sqrt{s})^{2}$ on $s$: as $s$ increases, the kinematically allowed $q^2$ region shrinks. In the finite-width treatment, the $b_1(1235)$ resonance is not approximated as a state with a fixed invariant mass; instead, its invariant mass $s$ varies across the $\omega\pi$ spectrum. Accounting for the finite width of the \(b_1(1235)\) resonance is essential for a reliable description of the $B_{\ell4}$ decay. Our results demonstrate that the finite-width effects of the $b_1(1235)$ resonance cannot be neglected in the four-body decay $B\to b_1(1235)(\to\omega\pi)e^+\nu_{e}$. After integrating over $q^2$ in the whole physical region, the corresponding decay widths can be obtained,
\begin{align}
\label{Ep:dGammaALL}
\Gamma(B\to b_1(1235)(\to &\omega\pi )e^+\nu_e)
 = (0.531^{+0.181}_{-0.149})\times 10^{-16}~{\rm GeV},
\nonumber\\
\Gamma(B \to b_1(1235)(\to &\omega\pi )\mu^+\nu_\mu)
 = (0.530^{+0.181}_{-0.148})\times 10^{-16}~{\rm GeV},
\nonumber\\
\Gamma(B \to b_1(1235)(\to &\omega\pi )\tau^+\nu_\tau)
 = (0.478^{+0.112}_{-0.096})\times 10^{-16}~{\rm GeV},
\end{align}
Finally, using the lifetime $\tau_{B^+} = (1.638\pm0.004)$ ps and $\tau_{B^0} = (1.512\pm0.003)$ ps issued by the PDG~\cite{ParticleDataGroup:2026mpi}, we have calculated the branching ratio of the $B_{\ell4}$ decay $B \to b_1(1235)(\to \omega\pi)\ell^+\nu_\ell$. To assess the impact of the finite-width effects, we also present the branching fraction for the three-body semileptonic decay $B\to b_1(1235)\ell^+\nu_\ell$ in Table~\ref{table:sitiBR}. For comparison, we include the corresponding theoretical predictions from LCSR (I)~\cite{Zeng:2025gft}, LCSR (II)~\cite{Sun:2011ssd}, and CLFQM~\cite{Kang:2018jzg}. In Table~\ref{table:sitiBR}, the branching ratios of $B\to b_1(1235)(\to \omega\pi)\ell^+\nu_\ell$ in the $e$ and $\mu$ channels are almost the same. This indicates that the lepton mass effect effects are negligible, consistent with the expectation of lepton-flavor universality. The branching ratios of the three-body semileptonic decay $B\to b_1(1235)\ell^+\nu_\ell$ in the $e$ and $\mu$ channels are close to those from LCSR(I)~\cite{Zeng:2025gft}. For the $\tau$ channel, the branching ratio is somewhat smaller than those in the other two channels. This is mainly because the larger $\tau$ mass reduces the available phase space. It is worth noting that, in both the electron and muon channels, the branching fractions for the four-body and three-body semileptonic decays differ by nearly a factor of three. This difference may partly arise from the finite width of the $b_1(1235)$ resonance. In the finite-width treatment, the $\omega\pi$ invariant mass squared $s$ is integrated over its kinematically allowed range, leading to a four-body decay width that can differ from the result obtained in the narrow-width approximation (NWA)~\cite{Cheng:2020mna}. Currently, no direct experimental measurement of this four-body process is available yet. Our results can provide a reference for future experimental studies.

For multibody decays, a cascade description in terms of sequential decay processes is commonly adopted within the factorization approximation~\cite{Cheng:2022vbw,Wu:2025dio,Cheng:2020iwk}. We include the decay $\omega \to \pi^+\pi^-\pi^0$. Under the NWA for the $\omega$ meson, the branching fraction for $B\to b_1(1235)(\to\omega (\to \pi^+\pi^-\pi^0)\pi)\ell^+\nu_\ell$ can then be predicted,
\begin{align}
&\mathcal{B}(B\to b_1(1235)(\to \omega(\to \pi^+\pi^-\pi^0)\pi )\ell^+ \nu_\ell)\nonumber \\
& = \mathcal{B}(B\to b_1(1235)(\to \omega\pi )\ell^+ \nu_\ell)\times \mathcal{B}(\omega \to \pi^+\pi^-\pi^0),
\end{align}
where the $\mathcal{B}(\omega \to \pi^+\pi^-\pi^0) = 0.892\pm0.007$~\cite{ParticleDataGroup:2026mpi} and
\begin{align}
\mathcal{B}(B^+\to b_1(1235)^0(\to \omega&(\to\pi^+\pi^-\pi^0)\pi^0)e^+ \nu_e)
\nonumber \\
& = 1.179^{+0.415}_{-0.337}\times 10^{-4},
\nonumber \\
\mathcal{B}(B^+\to b_1(1235)^0(\to \omega&(\to\pi^+\pi^-\pi^0)\pi^0)\mu^+ \nu_\mu)
\nonumber \\
& = 1.178^{+0.414}_{-0.337}\times 10^{-4},
\nonumber \\
\mathcal{B}(B^+\to b_1(1235)^0(\to \omega&(\to\pi^+\pi^-\pi^0)\pi^0)\tau^+ \nu_\tau)
\nonumber \\
& = 1.062^{+0.376}_{-0.307}\times 10^{-4},
\nonumber \\
\mathcal{B}(B^0\to b_1(1235)^-(\to \omega&(\to\pi^+\pi^-\pi^0)\pi^-)e^+ \nu_e)
\nonumber \\
& = 1.088^{+0.382}_{-0.311}\times 10^{-4},
\nonumber \\
\mathcal{B}(B^0\to b_1(1235)^-(\to \omega&(\to\pi^+\pi^-\pi^0)\pi^-)\mu^+ \nu_\mu)
\nonumber \\
& = 1.087^{+0.383}_{-0.311}\times 10^{-4},
\nonumber \\
\mathcal{B}(B^0\to b_1(1235)^-(\to \omega&(\to\pi^+\pi^-\pi^0)\pi^-)\tau^+ \nu_\tau)
\nonumber \\
& = 0.980^{+0.347}_{-0.284}\times 10^{-4}
\end{align}
The above results give the branching fractions for six six-body $B$-meson decay channels arising from the cascade decay involving $b_1(1235)\to\omega\pi$ and $\omega\to\pi^+\pi^-\pi^0$. These predictions provide a more complete description of the experimentally accessible final states and extend the phenomenological study of the corresponding multi-body semileptonic decays. They also provide useful theoretical benchmarks for future experimental searches and measurements. Future comparisons between these predictions and experimental data may help test the underlying weak transition dynamics and improve our understanding of the resonance effects in such multi-body decay processes.

\section{Summary}\label{sec:summary}

In this paper, we study the $B_{\ell4}$ decay $B\to b_1(1235)(\to \omega\pi)\ell^+ \nu_\ell$ using the QCD sum rules method. We treat the $b_1(1235)$ state as an intermediate resonance in this process and construct its transverse LCDA $\phi _{2;b_1(1235)}^\bot(x,\mu)$ of the $b_1(1235)$ resonance using the LCHO model. The model parameters are then determined by three additional constraints, its behavior is shown in Fig.~\ref{fig:DAb1}. Our model exhibits an axisymmetric distribution, which is consistent with the QCDSR~\cite{Yang:2007zt} predictions. These results can be used as nonperturbative inputs for studying the $B\to b_1(1235)\ell^+ \nu_\ell$ transition. Using the LCDA determined above, we calculate the TFFs at the maximum-recoil point $q^2 = 0$ within the LCSR framework, with the results presented in Table~\ref{table:A1A2V0}. We then extrapolate the TFFs over the full physical $q^2$ range to obtain their $q^2$ dependence, as shown in Fig.~\ref{fig:b1TFFs}. In particular, our prediction for the overall $q^2$ dependence of $V_1(q^2)$ agrees well with the LCSR(I) predictions~\cite{Zeng:2025gft}. Using the TFFs obtained above, we calculate the differential and total decay widths, as well as the branching fractions, for the $B_{\ell4}$ decay $B \to b_1(1235)(\to \omega\pi)\ell^+ \nu_\ell$. The $q^2$ dependence of the differential decay widths is shown in Fig.~\ref{fig:dGammadqb1omega}. In addition, Fig.~\ref{fig:dGammadqb1omega3D} presents the three-dimensional distribution of the double-differential decay width $d^2\Gamma/dsdq^2$ for $B \to b_1(1235)(\to \omega\pi)e^+\nu_e$ as a function of $s$ and $q^2$. The total decay width is obtained by integrating the differential decay width over the full physical $q^2$ range. Meanwhile, using the $B^+$ and $B^0$ lifetimes reported by the PDG~\cite{ParticleDataGroup:2026mpi}, we further predict the branching fractions for the four-body decay. To assess the impact of the finite width of the $b_1(1235)$ resonance, we also present the branching fractions for the three-body decay $B\to b_1(1235)\ell^+\nu_\ell$ within the NWA. The results are summarized in Table~\ref{table:sitiBR}. As shown in the table, the branching fractions for the four-body decay in the electron and muon channels are approximately $1/3$ times of those for the corresponding three-body decay. This indicates that finite-width effects have a substantial impact on the integrated decay rates. Since the $\omega$ can further decay into $\pi^+\pi^-\pi^0$, the branching ratio of $B \to b_1(1235)(\to\omega(\to\pi^+\pi^-\pi^0)\pi)\ell^+\nu_\ell$ is calculated under the NWA. This provides a theoretical reference for future experiments that observe the decay $B \to b_1(1235)(\to \omega(\to\pi^+\pi^-\pi^0)\pi)\ell^+ \nu_\ell$. Currently, no direct experimental measurement of this $B_{\ell4}$ process has been reported. It is hoped that in the near future, the results of this work can provide a theoretical reference for future experimental analyses of the $B\to b_1(1235)(\to \omega\pi)\ell^+ \nu_\ell$ final state. They can also provide a theoretical basis for further determining the internal structure of the $b_1(1235)$ resonance.

\section{Acknowledgments}

This work was supported by the National Natural Science Foundation of China under Grant No.12665017, No.12265010, the Project of Guizhou Provincial Department of Science and Technology under Grants No.MS[2025]219 and No.CXTD[2025]030.

\appendix
\section{}\label{appendixA}
The general analytic expression for the TFFs obtained by the LCSR method is given below:
\begin{align}
{\cal R}_i(q^2)&  =  {\cal N}_i \bigg[ A_i \phi_{2;{b_1(1235)}}^\bot(u,\mu) \! +  \! B_i\psi_{3;{b_1(1235)}}^\|(u)  \! -  \!  C_i\phi_{4;{b_1(1235)}}^\bot(u)
\nonumber\\
& - D_i I_L(u) - E_i H_3(u)  + G_i \Phi_{3;{b_1(1235)} }^\bot (\underline \alpha) + J_i \Psi_{4;{b_1(1235)}}^\bot (\underline \alpha)
\nonumber\\
&+ Q_i \widetilde \Psi_{4;{b_1(1235)}}^ \bot (\underline \alpha)\bigg], \label{Eq:VViTFFs}
\end{align}
where $i = 1, 2, 3$, corresponding to ${\cal R}_1(q^2) = V_1(q^2)$, ${\cal R}_2(q^2) = V_2(q^2)$ and ${\cal R}_3(q^2) = A(q^2)$. Whereas the function with $i = 0$ gives the combination ${\cal R}_0(q^2) =  V_3(q^2)-V_0(q^2)$. The auxiliary TFF $V_3(q^2)$ is defined as:
\begin{align}
V_3(q^2)  =  \frac{m_B+m_{b_1(1235)}}{2 m_{b_1(1235)}} V_1(q^2) - \frac{m_B-m_{b_1(1235)}}{2 m_{b_1(1235)}} V_2(q^2).
\end{align}
Where $V_3(q^2)$ is defined as a linear combination of $V_1(q^2)$ and $V_1(q^2)$, it is not an independent physical TFF. To ensure that the hadronic matrix element has no unphysical kinematic singularity at $q^2 = 0$, the condition $V_0(0) = V_3(0)$ must be satisfied. Thus, from the relation among $V_0(q^2)$, $V_3(q^2)$ and ${\cal R}_0(q^2)$, we have $V_0(q^2) = V_3(q^2)-{\cal R}_0(q^2)$.
The individual coefficients for the two-particle and three-particle distribution amplitudes are given below:
\begin{align}
{\cal N}_0& = \frac{m_b m_{b_1(1235)} f_{b_1(1235)}^\bot q^2}{2 f_B m_B^2}
\nonumber\\
{\cal N}_1& = \frac{m_b m_{b_1(1235)}^2 f_{b_1(1235)}^\bot}{f_B m_B^2(m_B+m_{b_1(1235)})},
\nonumber\\
{\cal N}_2& = \frac{m_b(m_B+m_{b_1(1235)}) m_{b_1(1235)}^2 f_{b_1(1235)}^\bot}{f_B m_B^2},
\\
A_0& = -\int_0^1\frac{du}{u}e^{\frac{m_B^2- s(u)}{M^2}}\bigg[ \frac{1}{m_{b_1(1235)}^2} \Theta (c(u,s_0))\bigg],
\nonumber\\
A_1& = \frac{\cal C}{u} \int_0^1\frac{du}{u}e^{\frac{m_B^2- s(u)}{M^2}}\bigg[ \frac{1}{m_{b_1(1235)}^2} \Theta (c(u,s_0))\bigg],
\nonumber\\
A_2& = \int_0^1\frac{du}{u}e^{\frac{m_B^2- s(u)}{M^2}}\bigg[ \frac{1}{m_{b_1(1235)}^2} \Theta (c(u,s_0))\bigg],
\\
B_0& = -\int_0^1\frac{du}{u}e^{\frac{m_B^2- s(u)}{M^2}}\bigg[ \frac{2-u}{u M^2} \widetilde \Theta (c(u,s_0))\bigg],
\nonumber\\
B_1& = m_{b_1}^2 \int_0^1\frac{du}{u}e^{\frac{m_B^2- s(u)}{M^2}}\bigg[ \frac{1}{m_{b_1(1235)}^2} \Theta (c(u,s_0))\bigg],
\nonumber\\
B_2& = -\int_0^1\frac{du}{u}e^{\frac{m_B^2- s(u)}{M^2}}\bigg[ \frac{1}{M^2} \widetilde \Theta (c(u,s_0))\bigg],
\\
C_0& = -\frac{1}{4}\int_0^1\frac{du}{u}e^{\frac{m_B^2- s(u)}{M^2}}\bigg[\frac{m_b^2}{u^2M^4} \widetilde{\widetilde\Theta}(c(u,s_0))+ \frac{\widetilde\Theta(c(u,s_0))}{uM^2} \bigg],
\nonumber\\
C_1& = \frac{1}{4}\int_0^1\frac{du}{u}\,e^{\frac{m_B^2- s(u)}{M^2}} \bigg[\frac{m_b^2{\cal C}}{u^3M^4}\widetilde{\widetilde\Theta}(c(u,s_0)) \nonumber\\
&+  \frac{{\cal C} - 2m_b^2}{u^2M^2}\widetilde\Theta(c(u,s_0))
- \frac{1}{u}\Theta(c(u,s_0))\bigg],
\nonumber\\
C_2& = \frac{1}{4}\int_0^1\frac{du}{u}e^{\frac{m_B^2- s(u)}{M^2}}\bigg[\frac{m_b^2}{u^2M^4} \widetilde{\widetilde\Theta}(c(u,s_0))+ \frac{\widetilde\Theta(c(u,s_0))}{uM^2} \bigg],
\\
D_0& = -\int_0^1\frac{du}{u}e^{\frac{m_B^2\!-\! s(u)}{M^2}}\!\bigg[(4-2u)\bigg[\frac{{\cal C}}{u^3 M^4}\widetilde{\widetilde\Theta}(c(u,s_0))
\nonumber\\
&-\frac{2\widetilde\Theta (c(u,s_0))}{u^2 M^2}\bigg]
+2\bigg(\frac{2 m_b^2\widetilde{\widetilde\Theta} (c(u,s_0))}{u^2 M^4}+\frac{1}{uM^2}
\widetilde{\widetilde\Theta}(c(u,s_0))\bigg)\bigg],
\nonumber\\
D_1& =  2\int_0^1\frac{du}{u}e^{\frac{m_B^2- s(u)}{M^2}}\bigg[\frac{\cal C}{u^2M^2}\widetilde\Theta(c(u,s_0)) -\frac{\Theta(c(u,s_0))}{u}\bigg],
\nonumber\\
D_2& = 2\int_0^1\frac{du}{u}e^{\frac{m_B^2\!-\! s(u)}{M^2}}\!\bigg[\frac{{\cal C}\! -\! 2m_b^2}{u^2 M^4}\widetilde{\widetilde\Theta}(c(u,s_0)) \!-\! \frac{\widetilde\Theta (c(u,s_0))}{uM^2}\bigg],
\\
E_0& =  \int_0^1\frac{du}{u}e^{\frac{m_B^2- s(u)}{M^2}}\bigg[ \frac{2-u}{u M^2} \widetilde \Theta (c(u,s_0))\bigg],
\nonumber\\
E_1& = \int_0^1\frac{du}{u}e^{\frac{m_B^2- s(u)}{M^2}}\bigg[ \frac{2m_b^2}{uM^2}\widetilde \Theta(c(u,s_0)) + \Theta(c(u,s_0))\bigg],
\nonumber\\
E_2& =  \int_0^1\frac{du}{u}e^{\frac{m_B^2- s(u)}{M^2}}\bigg[ \frac{1}{M^2} \widetilde \Theta (c(u,s_0))\bigg],
\\
G_0& = -\int {\cal D} \alpha_i\!\int_0^1 dv e^{\frac{m_B^2 \!- \!s(X)}{M^2}} \frac{1}{2X^2M^2}\! \Theta(c(X,s_0))\ 4v,
\nonumber\\
G_2& = \int {\cal D} \alpha_i\!\int_0^1 dv e^{\frac{m_B^2 \!- \!s(X)}{M^2}} \frac{1}{2X^2M^2}\! \Theta(c(X,s_0))\ 4v,
\\
J_0& = \int {\cal D} \alpha_i\!\int_0^1 dv e^{\frac{m_B^2 \!- \!s(X)}{M^2}} \frac{1}{2X^2M^2}\! \Theta(c(X,s_0))\ (4v-1),
\nonumber\\
J_1& = \int{\cal D}\alpha_i\int_0^1{dv}e^{\frac{m_B^2- s(X)}{M^2}}\bigg[\frac{\underline{\cal C}}{2X^3M^2}\nonumber\\
&-\frac{1}{2X^2}\bigg]\Theta(c(X,s_0))\ (4v-1),
\nonumber\\
J_2& = \int {\cal D} \alpha_i\!\int_0^1 dv e^{\frac{m_B^2 \!- \!s(X)}{M^2}} \frac{1}{2X^2M^2}\! \Theta(c(X,s_0))\ (4v-1),
\\
Q_0& =  -\int {\cal D} \alpha_i\!\int_0^1 dv e^{\frac{m_B^2 \!- \!s(X)}{M^2}} \frac{1}{2X^2M^2}\! \Theta(c(X,s_0)),
\nonumber\\
Q_1& = \int{\cal D}\alpha_i\int_0^1{dv}e^{\frac{m_B^2- s(X)}{M^2}}\bigg[\frac{\underline{\cal C}}{2X^3M^2}-\frac{1}{2X^2}\bigg] \Theta(c(X,s_0)),
\nonumber\\
Q_2& =  \int {\cal D} \alpha_i\!\int_0^1 dv e^{\frac{m_B^2 \!- \!s(X)}{M^2}} \frac{1}{2X^2M^2}\! \Theta(c(X,s_0)).
\end{align}
Here ${\cal N}_3 = {\cal N}_2$, $A_3 = A_2$, $C_3 = C_2$, and $B_3 = D_3 = E_3 = G_1 = G_3 = J_3 = Q_3 = 0$, $\mathcal{C}  =  m_b^2 + u^2 m_{b_1(1235)}^2 - q^2$, $s(u)  =  [m_b^2 - \bar{u}(q^2 - u m_{b_1(1235)}^2)] / u$ with $\bar{u}  =  1 - u$. We further introduce $X  =  a_1 - a_3$, and $c(u, s_0)  =  u s_0 - m_b^2 + \bar{u} q^2 - u \bar{u} m_{b_1(1235)}^2$. $\varTheta ( c( u,s_0 ))$ is the usual step function. $\widetilde{\varTheta }( c( u,s_0 ))$ and $\tilde{\tilde{\Theta}}\big(c(u,s_0)\big)$ are generalized step functions arising from partial integration, their explicit definitions can be found in Ref.~\cite{Fu:2014uea}. The simplified functions $I_L(u)$ and $H_3(u)$ are defined to Ref.~\cite{Cheng:2017bzz}.

\end{document}